\documentclass{jfm}
\usepackage[dvips]{graphicx}
\usepackage{psfrag}
\usepackage{amssymb,amssymb}
\usepackage{latexsym}    

\begin{document} 
\title[Long surface wave instability in dense 
granular flows]
{Long surface wave instability in dense granular flows.}

\author[Y. 
Forterre and O. Pouliquen] 
{Y\ls O\ls \"E\ls L\ns \ns F\ls O\ls R\ls T\ls E\ls R\ls R\ls E\ls 
\ and O\ls L\ls I\ls V\ls I\ls E\ls R\ns  P\ls O\ls
U\ls L\ls I\ls Q\ls U\ls E\ls N\ns }

\affiliation{IUSTI, UniversitŽ de Provence, CNRS UMR 6595,\\ 5 rue 
Enrico Fermi, 13453 Marseille cedex 13, France.} 

\date{2002} 

\maketitle

\begin{abstract}

In this paper we present an experimental study of the long surface 
wave instability that can develop when a granular material flows down 
a rough inclined plane. The threshold and the dispersion relation of 
the instability are precisely measured by imposing a controlled 
perturbation at the entrance  of the flow and measuring its evolution 
along the slope.  The results are compared with the prediction of a 
linear stability analysis conducted in the framework of the 
depth-averaged or Saint-Venant equations. We show that when the 
friction 
law proposed in Pouliquen (1999a)  is  introduced in the Saint-Venant 
equations, the theory is able to predict  quantitatively the 
stability 
threshold and the phase velocity of the  waves but fails in  
predicting the observed cutoff frequency. The instability is shown 
to be of the same nature as the long wave instability observed in 
classical fluids but with characteristics that can dramatically  
differ due to the specificity of the granular rheology.  

\end{abstract}

\section{Introduction}

Natural gravity flows such as mud flows or debris flows  
can be destructive.  Very often the material does not flow 
continuously along the slope of mountains, but develops in a 
succession of surges. These surges are of large amplitude and can be 
destructive as they are able to carry large debris. The breaking of 
an initially continuous flow to a succession of waves is often 
attributed to an inertial instability that can develop in thin liquid 
flows down slopes.   However, while this instability has been 
extensively studied for the case of Newtonian fluids, the formation 
of long surface waves in particulate flows has been much less 
investigated. The goal of this paper is to clarify the dynamics of 
the long-wave instability for a cohesionless granular material 
flowing 
down a rough inclined plane.  

Long-wave free surface instability 
in gravity flows is a phenomenon common to many fluids.  For a 
viscous 
fluid in the laminar regime, the instability is often called 
``Kapitza 
instability'' after the pioneering work by Kapitza $\&$ Kapitza 
(1949).  They have observed that a thin film of water flowing along a 
vertical wall does not remain uniform but deforms in a succession of 
transverse waves propagating down the wall.  Since Kaptiza's work, 
many 
experimental and theoretical studies have been devoted to thin 
viscous flows
 down inclined planes 
(see,  for instance, the review of 
\cite{chang94} or \cite{oron97}). Free surface waves are also 
observed in turbulent flows down slopes, where they evolve into a 
series 
of bores more 
or less periodic called ``roll 
waves'' (\cite{cornish34};  \cite{dressler49}; \cite{needham84}; 
\cite{kranenburg92}). These roll waves  are  of great importance in 
hydraulic for flows 
in open channels. Similar surges are also observed in flows 
using more complex fluids, e.g. mud flows, gravity currents, flows 
of particles suspension (see  \cite{simpson97}). 
 However,  few precise experimental studies can 
be 
found for these non Newtonian fluids.  

The spontaneous formation of 
free surface waves in fluids having very different rheological 
properties comes from the fact that  the instability mechanism does 
not depend on the precise fluid  characteristics (see the clear 
discussion by Smith (1990) in the context  of viscous film flows).   
When a fluid layer flows down an inclined plane, a small perturbation 
of the free surface 
propagates at first order with a phase velocity which is different 
than the mean fluid velocity. If the fluid could adjust 
instantaneously its velocity to the local thickness variation, the 
wave would propagate without amplification (kinematic waves,  
\cite{witham74}).   However, because of inertia, the fluid does not 
immediately adjust its velocity when the wave arrives.  This delay 
can give rise to a positive mass flux under the wave leading to the 
growth of the perturbation. 

From a theoretical point of view, many studies concern the case of 
laminar Newtonian flows.  In this case the linear stability analysis 
of a film flow can be derived based on the Orr-Sommerfeld equation 
(\cite{benjamin57}; \cite{yih63}).  Amplitude equations and precise 
non 
linear models have been also recently developed 
(\cite{chang94}; \cite{ruyer00}).  For non Newtonian complex fluids, 
exact 
three dimensional analysis are often not possible because of the lack 
of 
sufficient knowledge of the constitutive equations.  For those 
systems, a 
classical approach for studying the stability of thin flows is the 
shallow water description.  Mass and momentum balances are written in 
a depth averaged form (Saint-Venant equations, \cite{stvenant71}).  
In 
this framework, the rheological characteristics of the fluid are 
mainly taken into account into the expression of the shear stress 
between the flowing layer and the surface of the plane.  The shear 
stress is a 
viscous force for laminar Newtonian flows (\cite{shkadov67}), a 
turbulent friction given by a Chezy formula for turbulent flows 
(\cite{kranenburg92}), a Bingham stress for mud flows 
(\cite{liu94}).  
It is then possible to study the linear stability 
analysis of the flow and to predict the threshold and growth of the 
waves.

Although the free surface instability mechanism is the same 
for different fluids, the precise characteristics of the wave 
development  (threshold, growth rate,..) dramatically depend on the 
rheological  properties of the material.  A precise experimental 
study of the instability could then serve as a test for the 
rheological 
laws proposed for complex fluids.  This is the spirit of this paper 
on the wave formation in cohesionless 
granular materials. Formation of long waves in granular flows down 
inclined planes has been reported in previous studies 
(\cite{savage79}; \cite{davies90},  \cite{vallance94}; 
\cite{ancey97}; \cite{daerr01}).  However, to our knowledge no 
precise measurements of the instability properties  have been carried 
out.

The rheology of dry granular materials flowing in a dense regime 
is  still an open problem (\cite{rajchenbach00};  
\cite{pouliquen02a}).  However, for the flow of thin granular layers 
on inclined planes, a Saint-Venant description seems to be 
relevant.   
Savage and Hutter first proposed this approach for describing the 
motion of a granular mass down an inclined plane (\cite{savage89}).   
For the basal friction stress they used a solid friction law 
where the shear stress at the interface between the flowing material 
and the inclined plane is proportional to the normal stress, i.e. to 
the 
weight of the  material column above the base. Although this 
description is 
able to correctly predict the spreading of a mass on steep slopes and 
smooth surfaces (\cite{savage89}; \cite{naaim97}; \cite{gray99}; 
\cite{wieland99}),  
it does not capture the existence of steady uniform  flows 
when the surface is rough and the material sheared in the bulk.  
Recently, a generalization of the basal friction law has been 
proposed based on scaling properties observed in experiments 
(\cite{pouliquen99a}). The generalization gives a friction 
coefficient which 
is no longer constant but depends on the local velocity and 
thickness.  Using this law in Saint-Venant description it is possible 
to 
quantitatively predict the motion of a mass down a rough inclined 
plane 
from initiation to deposit (\cite{pouliquen02b}). In this paper we 
want to check to what extent the same approach using the generalized 
friction law can quantitatively predict the long wave instability for 
dry granular materials. 

To this end we have carried out precise 
measurement of the linear development of the instability in the same 
spirit as Liu {\it et al.} (1993) did  for liquid films. Liu {\it et 
al.} have developed a method to induce at the entrance of the liquid 
flow a perturbation whose amplitude and frequency can be controlled. 
The instability being convective, the observed waves downstream 
result from the amplification of the entrance perturbations. By 
following the evolution of the injected perturbation for different 
frequencies, Liu {\it et al.} were able to experimentally measure the 
dispersion relation and precisely determine the instability 
threshold. In this paper we follow the same procedure for granular 
flows: in the limited range of inclinations where steady uniform flows are observed, we impose an  external forcing to control the wave development.  

The paper is organized as follows. We first present 
preliminary experimental observations of the instability without 
forcing ($\S\,2$). We show that a simple visual analysis of the free 
surface instability is not sufficient to conclude about the 
instability mechanism and that more precise measurements are 
necessary. The experimental set up, the forcing method and the 
measurement procedure are described in $\S\,3$. The linear stability 
analysis based on Saint-Venant equations is presented in $\S\,4$.  
Results and comparison between experiments and theory are presented 
for glass beads in $\S\,5$ and for sand in $\S\,6$. Discussion and 
conclusion are given in $\S\,7$.    
 
\section{Preliminary observations}

Free surface waves in granular flows down a rough inclined plane have 
been reported previously by several authors.  Using glass beads, 
Savage (1979) and  Vallance (1994) observed the deformation of the 
free surface in 
a succession of  surges down a long chute flow.  Ancey (1997) 
reported formation of surge waves  with sand.  
Recently, Daerr (2001) has noticed an instability at the rear of 
avalanche 
fronts for flows of glass beads on velvet. All these experiments were 
performed on rough surfaces in a dense regime, i.e. with a mean 
volume fraction of the medium almost constant.  Waves have also been 
recently observed for granular flows on flat smooth surfaces 
(\cite{prasad00}; \cite{louge01}) 
but dramatic variations of  density are observed in this case, with 
very dilute regions. We will not discuss this case here.

\begin{figure} 
	\centering  
\includegraphics[scale=1]{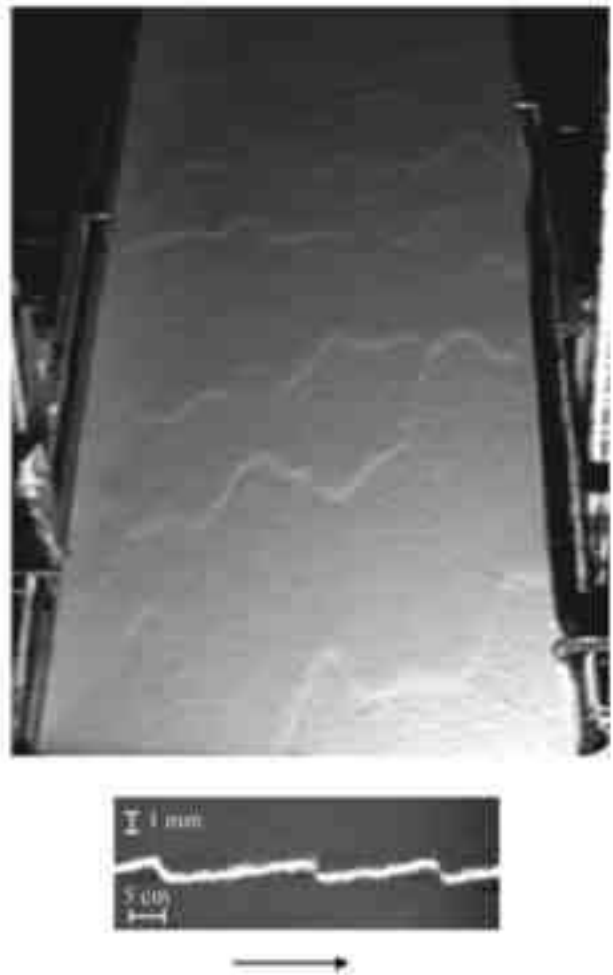} 
\caption{Free surface of the sand flow showing the long  surface 
waves. Material is 
flowing downwards; the plane width is 70 cm.  Inset: typical 
thickness 
profile of the natural waves measured with a laser sheet light (see 
\cite{pouliquen99a} for the description of the method). The arrow 
gives the flow direction. (sand 0.8 mm in  diameter, 
$\theta=34^{\circ}$, $h=$4.6 mm).} \label{figureondesable} 
\end{figure}

Figure \ref{figureondesable} shows the typical free 
surface we observe for a flow of sand 0.8 mm in diameter down an 
inclined plane made rough by gluing one layer of sand grains on it.  
A 
silo at the top of the plane (2 m long, 70 cm wide) continuously 
supplies the flow at a constant flow rate.  Waves first appear close 
to the outlet of the silo as two-dimensional deformations.  They 
rapidly amplify 
downstream and break down in the transverse direction due to 
secondary 
instabilities.  The shape of the saturated waves is highly non linear 
as shown by the typical thickness profile plotted in inset of figure 
\ref{figureondesable}. The distance between two surges is typically 
of the order of 20 cm, which is very large compared to the thickness 
of the layer 
(which is about 5 mm for the case of figure \ref{figureondesable}).

\begin{figure}
\centering

 \includegraphics[scale=0.55]{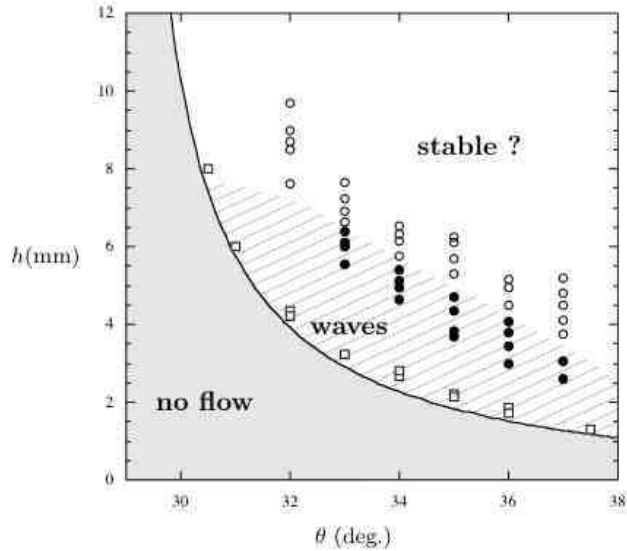}
\caption{Qualitative stability diagram for sand 0.8 mm in diameter. 
The 
hatched zone is the domain where waves are observed. ($\bullet$): 
waves, ($\circ$): no wave, ({\footnotesize $\Box$}): flow threshold.}
\label{figuresablephase}
\end{figure}

The observed pattern does look similar to patterns observed in liquid 
flows.  However, a major difference appears when changing the 
experimental conditions, i.e.  the thickness of the layer and the 
inclination of the plane.  Figure \ref{figuresablephase} shows a 
qualitative phase diagram obtained simply by visual observations of 
the free surface.  The striking result of this preliminary analysis 
with sand is that the waves are observed mainly close to the 
threshold 
of the flow (low inclinations and thin flows).  In this region, large 
amplitude waves are observed.  When one increases the thickness of 
the 
layer at a fixed inclination, the waves form further and further 
downstream and eventually disappear for thick flows i.e.  for more 
rapid flows.  Naively, one would say that this observation is in 
contradiction with the explanation in terms of an inertial 
instability, which should develop preferentially in rapid flows 
rather 
than in slow flows.


Another striking preliminary observation is made using glass beads as 
a granular material.  In this case, no wave was observed in our set-up,
 whatever the thickness and the inclination.  The absence of 
noticeable deformation of the free surface is what made 
systematic measurements of the steady uniform regime  possible and 
allowed to put in 
evidence flow scaling laws (\cite{pouliquen99a}).  This point is also 
intriguing as one would expect the long-wave instability to occur as 
in other fluids for fast enough flows (\cite{vardoulakis02}).


The situation is then rather confused and the preliminary 
observations 
raise several questions.  First, why are the flow of sand and the 
flow of 
glass beads so different?  Secondly, in sand the waves seem to 
develop for slow flows and disappear for rapid flow in apparent 
contradiction with an inertial instability.  Are the waves due to a 
different instability mechanism?  Finally, why is no wave observed 
with glass beads whereas for rapid flows one would expect the 
instability to develop?  In order to properly answer those questions 
one need to precisely study the stability of the flow not only by 
visual observation of the free surface deformation but by careful 
measurement of the amplification of an initially imposed 
perturbation.  
This is the reason why, following Liu and Gollub's procedure (Liu {\it et al.} 1993), 
we have developed a forcing method.

\section{Experimental setup}

\begin{figure}
\centering

 \includegraphics[scale=0.65]{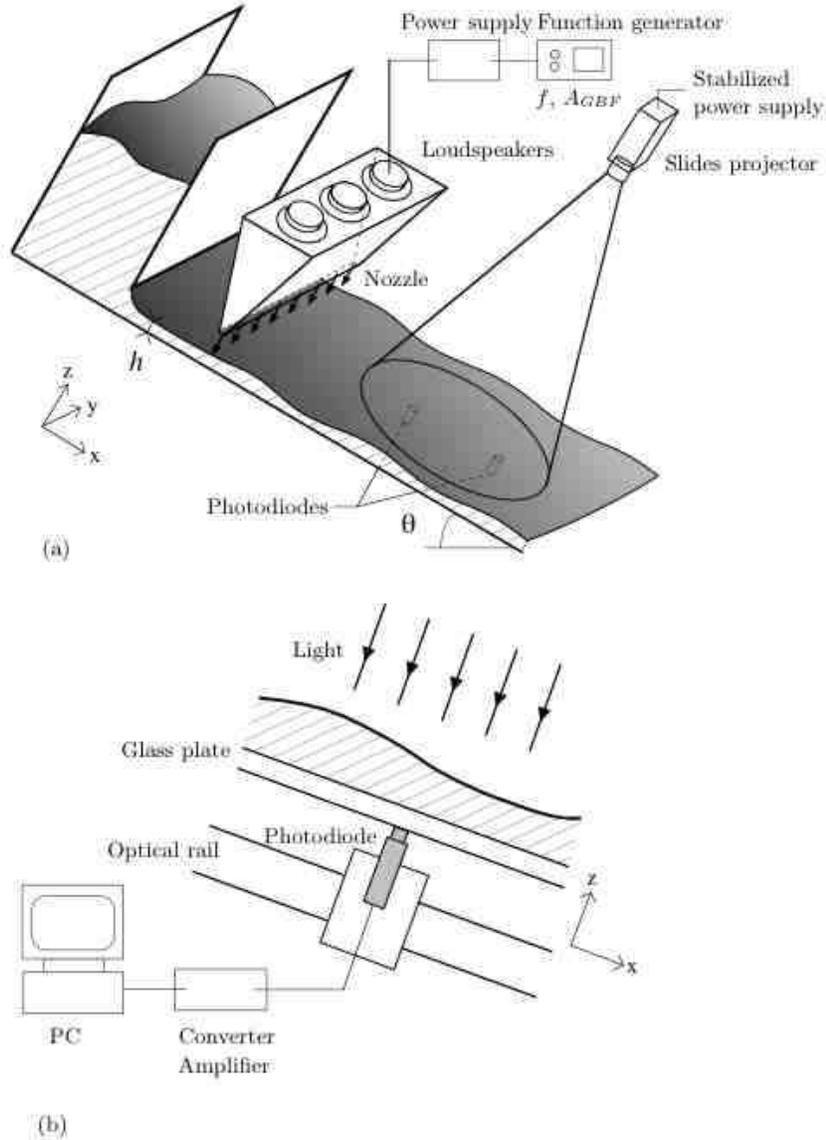}
\caption{(a) Sketch of the experimental set-up used to force the 
instability.
 (b) Description of the light detection method.}
\label{figuredispexp}
\end{figure}

\begin{figure}
\centering
 \includegraphics[scale=0.7]{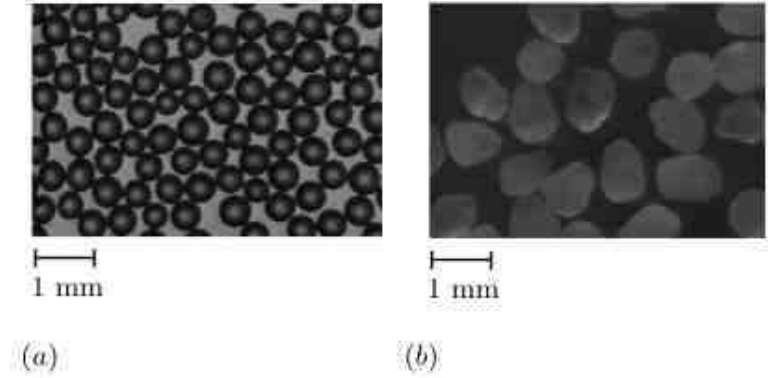}
\caption{Granular materials used. ($a$) Glass beads $d=0.5\pm0.05$ mm 
in 
mean diameter (grinding glass beads supplied by Potters-Ballotini, 
France).
  ($b$) Sand $d=0.8\pm0.1$ mm in diameter (sand of the Loire supplied 
by Sifraco, France).}
\label{figurematerial}
\end{figure}

The experimental set-up is presented in figure \ref{figuredispexp}.  
The plane is 2 m long and 35 cm wide.  The bottom plate as well as 
the 
side walls are glass plates.  The roughness is made by gluing on the 
bottom plate one layer of the particle used for the flow.  The flow 
rate is controlled by the opening of the silo.  In this paper we have 
used two granular materials: glass beads $d=0.5$ mm in mean diameter 
(the same as used in \cite{pouliquen99a}), and sand $d=0.8$ mm in 
mean 
diameter (figure \ref{figurematerial}).

\subsection{Forcing method}
 
\begin{figure}
\centering
 \includegraphics[scale=0.5]{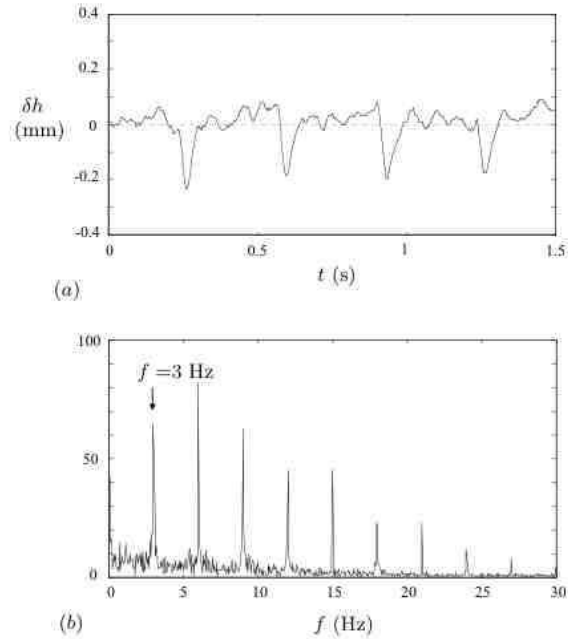}
\caption{Free surface deformations of the flow below the nozzle 
($f=$3 Hz, $A_{GBF}=1.5$ V). ($a$) Thickness variations. ($b$) 
Fourier 
transform of the thickness variations, $\vert a_{f}\vert$, averaged over 60 cycles 
(arbitrary units).}
\label{figurerepbuse}
\end{figure}

In order to impose a perturbation whose frequency and amplitude can 
be 
easily controlled we periodically blow a thin air jet on the free 
surface.  The jet is created by three loudspeakers embedded into a 
two 
dimensional nozzle with a 1 mm slit (figure \ref{figuredispexp}$a$).  
A 
homogeneous and localized air jet is then created, with an amplitude 
and a frequency controlled by the amplitude $A_{GBF}$ and frequency 
$f$ sent to the loudspeakers via a low frequency function generator.  
The nozzle is placed 30 cm from the outlet of the reservoir.  The 
typical free surface deformation we achieved with this set up is of 
the order of 0.25 mm and frequency varies between 1 Hz and 20 Hz.  A 
typical response of the free surface to the forcing air jet is 
plotted 
in figure \ref{figurerepbuse}($a$). In this figure, the measurement is carried out 
just below the nozzle and the signal sent to the loudspeaker is 
sinusoidal at a frequency of 3 Hz.  One observes that the thickness 
variation follows the forcing frequency but that the response is not 
sinusoidal but highly asymmetric.  When air is blown, the air jet is 
localized and the induced deformation important.  When the nozzle 
sucks the air, there is no influence on the free surface. This 
difference between localized ejection and diffuse injection is 
well-known and is used, for instance, for propulsion in water.  The non 
sinusoidal response clearly appears on the Fourier transform of the 
signal in figure \ref{figurerepbuse}($b$).  The perturbation is periodic at 
the imposed frequency, but many harmonics are also present with 
amplitude that can be higher than the fundamental.  This effect is 
more pronounced when the forcing frequency is low.  With this forcing 
method it is then difficult to inject low frequency modes at 
measurable amplitude, without injecting very high harmonics, which 
will interact in a non linear regime. Consequently,  no measurement 
of the linear 
evolution was possible below 1 Hz.


\subsection{Thickness measurement}

\begin{figure}
	\centering
\includegraphics[scale=0.6]{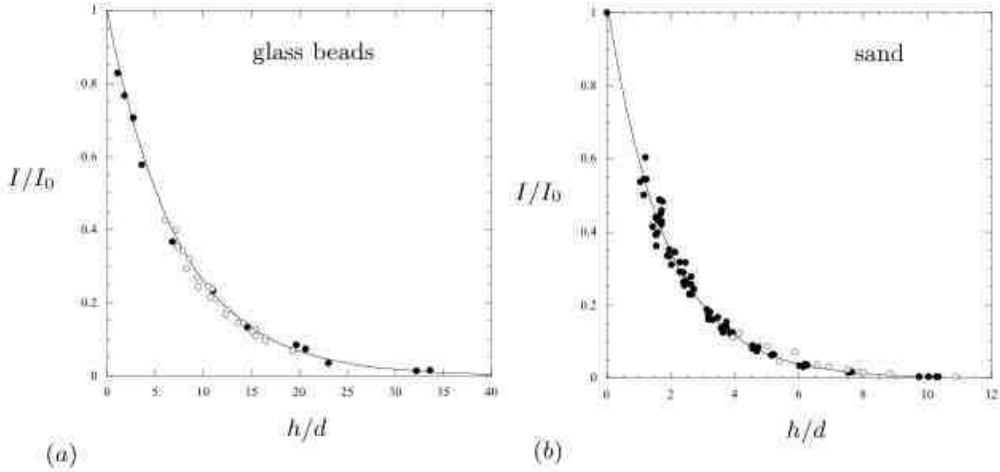}
    \caption{Calibration curve for the light transmitted through a 
    layer of grain as a function of the thickness $h$ of the layer, 
    ($a$) 
     glass beads $d=0.5$ mm in diameter,  ($b$) sand $d=0.8$ mm in 
diameter.
    ($\bullet$): static layers. ($\circ$): flowing layers. The solid 
    curve is the best exponential fit $I/I_{0} = \exp (-h/L_a)$ with 
      $L_a=7.40\,d$ for glass beads and $L_a=1.82\,d$ for sand.}
\label{figurecaliblumiere}
\end{figure}

In order to measure precisely the free surface deformation we use a 
light absorption method.  A parallel beam lights the plane from above 
as sketched in figure \ref{figuredispexp}($b$).  Two photodetectors 
are placed at different distances from the nozzle below the plane.  
The light measured by the detector varies exponentially with the 
thickness of the granular layer as shown by the calibration curves in 
figure \ref{figurecaliblumiere}.  The light intensity received by the 
detector is plotted versus the thickness (see \cite{pouliquen99a} for 
the thickness measurement method).  The data are well fitted by 
$I=I_{0}\exp(-h/L_a)$, the attenuation length being larger for the 
beads 
$L_a$=7.4 $d$, than for the sand $L_a$=1.82 $d$.  This calibration has 
been carried out for both  uniform layer at rest (black dots) or 
flowing layers (circles).  We observe that the two data sets 
collapse.  
Since the light absorption is a function of both the layer thickness 
and the volume fraction, this collapse means that the variations of 
the volume fraction 
 are negligible in the dense flow regime studied here.

The signal measured by the photodetector is digitalized by a PC 
trough 
an acquisition board at 100 kHz.  Notice that from the temporal 
evolution of the light amplitude, $I(t)$, it is possible without any 
calibration to get the deformation of the free surface $\delta 
h(t)=h(t)-\langle h(t) \rangle$ where $\langle h(t) \rangle$ is the 
mean thickness.  The exponential attenuation law leads to the 
following expression: $\delta h(t)=-L_a(\ln I(t)-\langle \ln 
I(t)\rangle) 
$, which is independent of $I_{0}$ and of the gain of the 
photodetectors.  With this method the deformation of the free surface 
is determined with a precision of about 0.05 mm.


\subsection{Measurement of the dispersion relation}

From the thickness measurement at two different points it is in 
principle possible to measure the dispersion relation.  To this end a 
perturbation at a given frequency $f$ is imposed to the flow by our 
forcing device.  The two photodetectors then give the free surface 
deformations $\delta h(x_{1},t)$ and $\delta h(x_{2},t)$ at two 
different locations $x_{1}$ and $x_{2}$.  By computing the Fourier 
transform of the deformations, one obtains at the two locations the 
amplitudes $a_{f}(x_{1})$ and $a_{f}(x_{2})$ and the phases 
$\phi_{f}(x_{1})$ and $\phi_{f}(x_{2})$ of the fundamental mode at 
frequency $f$ defined by : $\delta h_{f}(x,t)=a_{f}(x) \cos (2\pi f t 
- 
\phi_{f}(x))$.  If the evolution between $x_{1}$ and $x_{2}$ is 
assumed to be exponential, i.e.  in the linear regime of the 
instability, the deformation of the mode $f$ is supposed to vary as: 
$\delta h_{f}=a_{0}e^{\sigma x} \cos (2\pi f (t-x/v_{\phi}))$.  The 
spatial growth rate $\sigma(f)$ and the phase velocity $v_{\phi}(f)$ 
are then given by:

\begin{eqnarray} 
\sigma(f) 
=\frac{1}{x_{2}-x_{1}}\ln 
\left(\frac{a_{f}(x_{2})}{a_{f}(x_{1})}\right),\\
v_{\phi}(f)=\frac{2 \pi  (x_{2}-x_{1}) 
f}{\phi_{f}(x_{2})-\phi_{f}(x_{1})}.
\end{eqnarray}

The dispersion relation can then in principle be determined.  
Experimentally the difficulty is to determine a region of linear 
instability where 
the wave grows exponentially.  The amplitude of the deformation has 
to be small enough to remain in the linear evolution but not too 
small 
to be measured.
\section{Linear stability analysis}

In this section, we present the theoretical analysis of the long-wave 
instability for granular flows in the framework of the depth-averaged 
equations.  A previous analysis has been carried out by Savage (1989) 
using a Chezy formula to model the basal shear stress.  Here we 
present the theoretical prediction when the interaction between the 
granular layer and the rough plane is described by the empirical 
friction law derived from experimental measurements on steady uniform 
flows (\cite{pouliquen99a}).

\subsection{Depth-averaged equations}

\begin{figure}
\centering
 \includegraphics[scale=0.5]{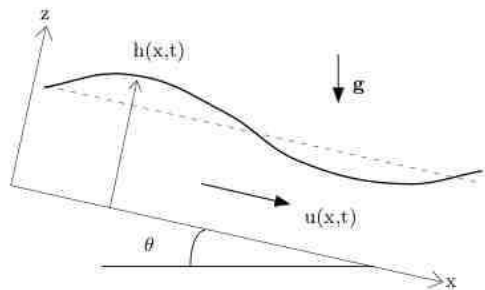}
\caption{Sketch of the flow.}
\label{figureanalyselineaire}
\end{figure}

A detailed derivation of depth-averaged equations in the context of 
granular flows can be found in Savage $\&$ Hutter (1989).  Assuming 
that the flow is incompressible and that the spatial variation of the 
flow takes place on a large scale compared to the thickness of the 
flow, we obtain the depth-averaged mass and momentum equations by 
integrating the full three-dimensional equations.  For a 
two-dimensional flow down a slope making an angle $\theta$ with the 
horizontal (see figure \ref{figureanalyselineaire}), depth-averaged 
equations reduce to:

\begin{eqnarray} \frac{\partial h}{\partial t} 
+\frac{\partial h u}{\partial x} & = &  0,\label{eqmoyintrom}\\ \rho 
\left( \frac{\partial hu}{\partial t} +\alpha \frac{\partial h  
u^2}{\partial x}\right) & = & \left(\tan \theta - \mu(u,h) -  
K\frac{\partial h}{\partial x}\right) \rho g h \cos \theta, 
\label{eqmoyintrop} \end{eqnarray}
where $h$ is the local thickness 
of the flow and $u$ is the averaged  velocity defined by $u=Q/h$, $Q$ 
being the flow rate per unit of  width.

The first equation (\ref {eqmoyintrom}) is the mass conservation.  
The 
second equation (\ref {eqmoyintrop}) is the momentum equation where 
the acceleration is balanced by three forces.  In the acceleration 
term, the coefficient $\alpha$ is related to the assumed velocity 
profile across the layer and is of order 1.  We will discuss the role 
of the parameter $\alpha$ in $\S\,5$.  The first force in the 
right-hand side is the gravity parallel to the plane.  The second 
term 
is the tangential stress between the fixed bottom and the flowing 
layer; it is written as a friction coefficient $\mu$ multiplied by 
the 
vertical stress $\rho g h \cos\theta$.  The friction coefficient 
is assumed to depend only on the local thickness $h$ and the local velocity $u$. This is a generalization of Savage $\&$ Hutter (1989)'s assumption of a constant solid friction. However, the friction coefficient $\mu$ could depend on other quantities such as the normal stress or the derivatives of $h$ and $u$.    The 
last term in (\ref {eqmoyintrop}) is a pressure force related to the 
thickness gradient.  The coefficient $K$ represents the ratio of the 
normal stress in the horizontal direction ($x$-direction) to the 
normal stress in the vertical direction ($z$-direction).  Recent 
numerical simulations have shown that for dense granular flows the 
horizontal and the vertical normal stresses are almost equal 
(Prochnow 
{\it et al.} 2000, Ertas {\it et al.} 2001). Therefore, we will take 
$K=1$ in the following. However, it should be noticed that for non 
uniform flows, the factor $K$ could depend on the divergence of the 
flow (\cite{gray99}; \cite{wieland99}).  

The great advantage of the Saint-Venant equations is that the 
dynamics 
of the flowing layer can be predicted without knowing in details the 
internal structure of the flow, although some information of the flow 
dynamics is lost. The complex and unknown 
three-dimensional rheology of the material is taken into account only 
in the 
friction term $\mu(u,h)$.  Moreover, this friction is easily 
determined experimentally by studying steady uniform flows.  These 
flows simply result from the balance between gravity and friction, 
i.e.  $\mu(u,h) = \tan \theta$.  This means that measuring how the 
mean velocity $u$ varies with the thickness $h$ and inclination 
$\theta$ is sufficient to determine the friction law.  Knowing the 
function $u(h,\theta)$ is then equivalent to knowing the friction law 
(see \cite{pouliquen99a}).\\


\subsection{Stability analysis}

We study here the stability of a steady uniform flow of thickness 
$h_{0}$ and averaged velocity $u_{0}$.  For the sake of simplicity, 
we 
first re-write the Saint-Venant equations (\ref{eqmoyintrom}) and 
(\ref{eqmoyintrop}) with the dimensionless variables given by: $ 
\tilde{h} = h/h_{0}$, $\tilde{u} = u/u_{0}$, $\tilde{x} = x/h_{0}$ 
and 
$\tilde{t} = (u_{0}/h_{0})t$.  We then obtain:
 \begin{eqnarray}
\frac{\partial \tilde{h}}{\partial \tilde{t}} +\frac{\partial \tilde{ 
h}\tilde{ u}}{\partial \tilde{ x}} & = & 0,\label{eqmoyadimm}\\
{F}^2 \left( \frac{\partial \tilde{ h}\tilde{u}}{\partial \tilde{ t}} 
+\alpha \frac{\partial \tilde{ h} {\tilde{ u}}^2}{\partial \tilde{ 
x}}\right) & = & \left(\tan \theta - \mu(\tilde{u},\tilde{h}) - 
\frac{\partial \tilde{h}}{\partial \tilde{x}}\right) \tilde{ h},
\label{eqmoyadimp}
\end{eqnarray}
where $F$ is the Froude number defined by:
\begin{equation}
	F =\frac{u_{0}} {\sqrt{gh_{0}\,\cos \theta}}.
	\label{eqdeffroude}
\end{equation}
The two dimensionless control parameters of the problem are therefore 
the Froude number $F$ and the angle of inclination $\theta$.  \\
We look for steady uniform solutions given by:
 \begin{equation}
\tilde{h}(\tilde{x},\tilde{t}) = 1, \; \tilde{u}(\tilde{x},\tilde{t}) 
= 1.
\label{eqalesu}
\end{equation}
For this basic state, the mass conservation (\ref{eqmoyadimm}) is 
satisfied and the momentum equation (\ref{eqmoyadimp}) reduces to the 
balance between gravity and friction: $\mu (1,1) = \tan \theta.$ The 
next step is to study the stability of (\ref{eqalesu}) by perturbing 
the flow: $ \tilde{h}(\tilde{x},\tilde{t}) = 1 + 
h_{1}(\tilde{x},\tilde{t})$, $\tilde{u}(\tilde{x},\tilde{t}) = 
1+u_{1}(\tilde{x},\tilde{t})$ with ($ 
h_{1}(\tilde{x},\tilde{t})\ll 1$, $u_{1}(\tilde{x},\tilde{t})\ll 1$) 
and by linearizing the 
depth-averaged equations which become:
\begin{eqnarray}
\frac{\partial h_{1}}{\partial \tilde{t}} + \frac{\partial 
h_{1}}{\partial \tilde{x}} 
+ \frac{\partial u_{1}}{\partial \tilde{x}} & = & 0,\label{eqlinm}\\
{F}^{2}\left( \frac{\partial u_{1} }{\partial \tilde{t}} + (\alpha 
-1)\frac{\partial h_{1}}{\partial \tilde{x}} + (2\alpha - 
1)\frac{\partial 
u_{1}}{\partial \tilde{x}} \right) & = & - a u_{1} - b h_{1} 
-\frac{\partial 
h_{1}}{\partial \tilde{x}},
\label{eqlinp}
\end{eqnarray}
where the dimensionless variables $a$ and $b$ are related to the 
friction law $\mu (\tilde{u},\tilde{h})$ by:
\begin{equation}
	a = {\left( \frac{\partial \mu}{\partial \tilde{u}}\right)}_{0}, \; 
b = 
	{\left( \frac{\partial \mu}{\partial \tilde{h}}\right)}_{0},
	\label{eqdefab}
\end{equation}
(the index `0' means that the derivatives are calculated for the 
basic 
state).  Note that to derive (\ref{eqlinp}), we have used the mass 
conservation (\ref{eqlinm}).

We then seek normal mode solutions for the perturbations: 
$h_{1}(\tilde{x},\tilde{t}) = \hat{h}\exp 
i(\tilde{k}\tilde{x}-\tilde{\omega} \tilde{t})$ and 
$u_{1}(\tilde{x},\tilde{t}) = 
	\hat{u}\exp i(\tilde{k}\tilde{x}-\tilde{\omega} \tilde{t})$ where 
$\tilde{k}$ is the dimensionless wavenumber and $\tilde{\omega}$ is 
the dimensionless pulsation, which,  once introduced in the 
linearized 
	equations (\ref{eqlinm}) and (\ref{eqlinp}), gives the following  
dispersion relation: 
 \begin{equation}
	-\tilde{\omega}^2 + 2\alpha \tilde{\omega}\tilde{k} + 
\frac{i}{F^2}\left( (a-b)\tilde{k}-a\tilde{\omega} 
	\right) + \left( \frac{1}{F^2}-\alpha \right) \tilde{k}^2= 0.
	\label{eqreldisp}
\end{equation}  
From this dispersion relation, we can easily compute the spatial 
growth rate and the phase velocity of the waves for a given imposed 
real pulsation.  Computation of the spatial stability analysis is 
given in Appendix A. This analysis shows that the flow is unstable 
when:
\begin{equation}
1-\frac{b}{a}> \alpha + \sqrt{\alpha(\alpha - 1) + \frac{1}{F^2}}.
\label{eqseuilspa}
\end{equation}

The above stability criterion is expressed as a function of $a$ and 
$b$ written in terms of the derivatives of the friction law 
$\mu(u,h)$.  Experimentally, we have access to the relation 
$u(h,\theta)$ for steady uniform flows.  It is therefore useful to 
write a stability criterion using the relation $u(h,\theta)$ instead 
of $\mu(u,h)$.  Moreover, we will see that this formulation gives an 
interpretation of the long-wave instability in terms of waves 
interactions.

As for steady uniform flows we have  $\mu(u,h) = \tan(\theta(u,h))$, 
it is easy to show that:
\begin{equation}
a  =  \frac{1}{\cos^2 \theta }{\left( \frac{\partial 
\theta}{\partial \tilde{u}}\right)}_{0}, \mbox{ and} \;  
\frac{b}{a}  =  -{\left( \frac{\partial \tilde{u}}{\partial 
\tilde{h}}\right)}_{0}.
\label{eqabu}
\end{equation}

The stability criterion (\ref{eqseuilspa}) may therefore be written 
as:
\begin{equation}
	\tilde{c}_{0}>\tilde{c}_{+},
	\label{eqseuilcine}
\end{equation}
where
\begin{eqnarray}
\tilde{c}_{0} & = & 1+\left( \frac{\partial \tilde{u}}{\partial 
\tilde{h}}\right)_{0},\label{eqdefco}\\
\tilde{c}_{+} & = & \alpha + \sqrt{\alpha(\alpha - 1) + 
\frac{1}{F^2}}.\label{eqdefcp}
\end{eqnarray}
It can be shown that $\tilde{c}_{0}$ is the (dimensionless) velocity 
of the 
kinematic waves, whereas $\tilde{c}_{+}$ is the (dimensionless) 
velocity of 
the ``gravity waves'' that propagate downstream (see Appendix B).  
The 
stability criterion (\ref{eqseuilcine}) therefore means that the flow 
is unstable when the velocity of the kinematic waves is larger than 
the velocity of the gravity waves.  Using this formulation, one can 
then conclude about the stability of the flow just from the velocity 
law $u(h,\theta)$ of steady uniform flows.


 \begin{figure}
  \centering 
 	 \includegraphics[scale=0.57]{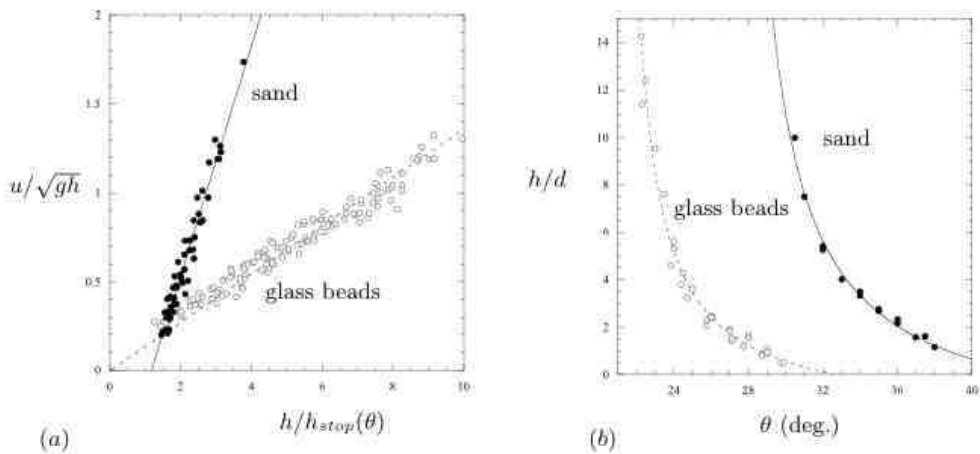} 
 	 \caption{($a$) Measurements of the averaged velocity of steady uniform 
 	 flows $u$ as a function of the thickness $h$ and the angle of 
 	 inclination
 	 $\theta$ for glass beads $d=$0.5 mm ($\circ$) (\cite{pouliquen99a}) 
 	 and sand $d=$0.8 mm ($\bullet$). The data collapse on a 
 	 curve when  $u/\sqrt{gh}$ is plotted as a 
 	 function of  
 	 $h/h_{stop}(\theta)$. ($b$) deposit function $h_{stop}(\theta)$ for 
 	 the glass beads ($\circ$)
 	 and the sand ($\bullet$).  The solid line is 
an interpolation given by 
$h_{stop}(\theta)=L((\tan\delta_{2} - \tan\delta_{1})/(\tan\theta - 
    \tan\delta_{1})-1)$ where  ($L/d=1.65$, $\delta_{1}=20.90^{\circ}$,  
    $\delta_{2}=32.76^{\circ}$) for the glass beads and ($L/d=2.03$, 
    $\delta_{1}=27.0^{\circ}$, 
    $\delta_{2}=43.4^{\circ}$) for the sand.  } 
 \label{figurerheologie}
 \end{figure}

\subsection{Velocity law for granular materials}

In this study we use two different granular materials: glass beads 
and 
sand. The velocity law for the glass beads has been previously 
measured (\cite{pouliquen99a}).
The mean velocity 
$u$, the inclination $\theta$ and the thickness $h$ are related 
through the following relation:

\begin{equation}
	\frac{u}{\sqrt{gh}} = \beta \frac{h}{h_{stop}(\theta)}.
	\label{equbille}
\end{equation}
where $\beta = 0.136$.  The function $h_{stop}(\theta)$ is the 
thickness of the deposit left by a steady uniform flow at the 
inclination $\theta$ (\cite{pouliquen96}; \cite{daerr99}).

In order to know the velocity law for the 0.8 mm sand we have also 
performed the same systematic measurements of the mean velocity as a 
function of thickness and inclination.  We observe that, as for glass 
beads, the velocity is correlated to the deposit function 
$h_{stop}(\theta)$ as shown in figure \ref{figurerheologie}.  
However, 
the master curve is different for sand and glass beads.

We can therefore write the velocity law in the cases of glass beads 
and 
sand in a similar form:

\begin{equation}
	\frac{u}{\sqrt{gh}} = -\gamma + \beta \frac{h}{h_{stop}(\theta)},
	\label{eqrheologiegranulairegenerale}
 \end{equation}
with
$$
 \left\{ \begin{array}{lllllll}
             \gamma   & = & 0 & \beta & = & 0.136 & \mbox{for glass 
beads},  
\\
    \gamma   & = & 0.77 & \beta & = & 0.65 & \mbox{for sand},
    \end{array}
    \right.
    \label{def:transport}
 $$
(the function $h_{stop}(\theta)$ for the glass beads and the sand 
is given in the caption of figure \ref{figurerheologie}).

From this relation we can then apply the stability criterion 
(\ref{eqseuilcine}) which compares the kinematic wave velocity and 
the 
gravity wave velocity. To compute the velocities, one needs to know 
the coefficient $a$ and $b$ given by the (\ref{eqabu}). From 
(\ref{eqrheologiegranulairegenerale}), one obtains:
\begin{equation}
a  = -\frac{h_{stop}(\theta)}{{\cos}^{3/2}\theta 
	{h_{stop}}'(\theta)}\frac{F}{(F\sqrt{\cos\theta}+\gamma)}, \mbox{ and} \;  
\frac{b}{a}  =  -\left( \frac{3}{2} + \frac{\gamma}{F\sqrt{\cos\theta}}\right).
\label{eqabgranul}
\end{equation}
The (dimensionless) velocity of the kinematic 
waves given by (\ref{eqdefco}) is therefore:
\begin{equation}
 \tilde{c}_{0}= \frac{5}{2} +\frac{\gamma}{F\sqrt{\cos\theta}}.
\label{eqc0granul}
\end{equation}
The speed of the gravity waves depends 
on the parameter $\alpha$ (see \ref{eqdefcp}).  With $\alpha=1$, the 
instability condition takes a particularly simple expression:
 \begin{equation} F >\frac{2}{3}\left( 1- 
\frac{\gamma}{\sqrt{\cos\theta}}\right).  \label{edfrcbille} 
\end{equation} The linear stability analysis predicts therefore an 
instability for granular flows when the Froude number is above a 
critical Froude number $F_{c}$.  For glass beads ($\gamma=0$) the 
critical Froude number is independent of the angle of inclination and 
given by $F_{c}=2/3$.  In the next section, we will compare these 
theoretical predictions to the experimental studies.

\section{Results for glass beads}

We first present the experimental results for the flow of glass 
beads.  
When there is no forcing, no wave was observed.  By imposing a 
forcing at the 
entrance of the flow we show that the instability exists but that the 
spatial growth rates are small.  Our inclined plane was therefore too 
short to allow the observation of the natural instability.

\subsection{Spatial evolution of a forcing wave: existence of a 
linear 
regime}

\begin{figure}
	\centering
\includegraphics[scale=0.55]{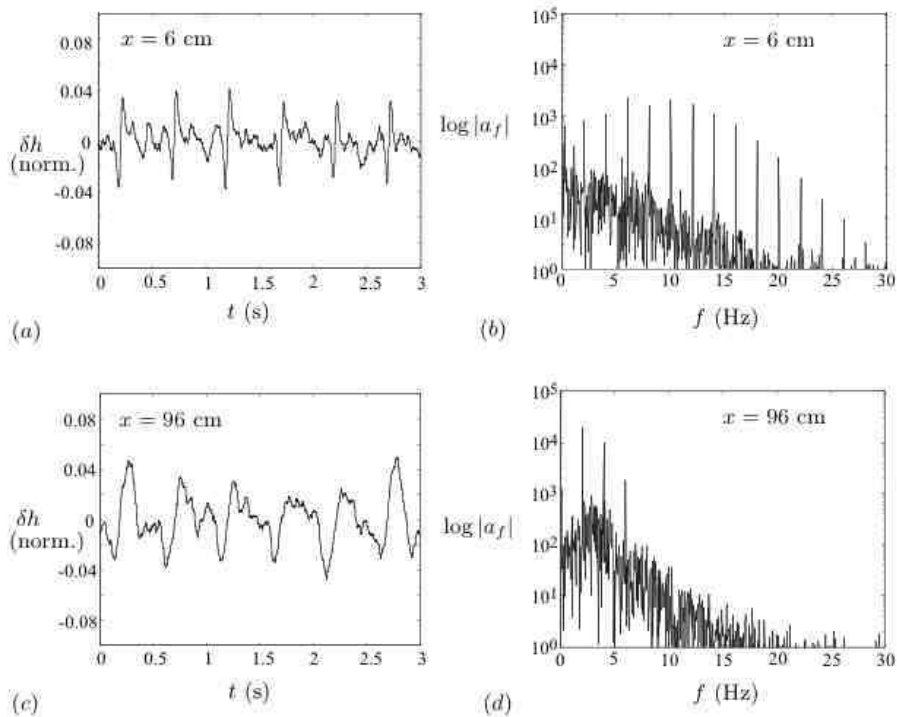}
    \caption{Wave measurement at two distances from the 
    nozzle,  
    $x=6$ cm and $x=96$ cm, in the unstable regime. ($f=2$ Hz, 
$\theta = 
29^{\circ}$, $h=$ 5.3 mm). ($a$) and ($c$) Local thickness 
variations. 
($b$) and ($d$) Power spectra averaged on  40 cycles.}
\label{figureevospabille}
\end{figure}

An example of an amplification of the imposed perturbation is shown 
in 
figure \ref{figureevospabille}.  The typical temporal signals 
measured 
at two locations in the unstable regime ($\theta = 29^{\circ}$, $h=$ 
5.3 mm) is shown for a forcing frequency $f=$2 Hz.  First, we note 
that the instability is convective, i.e.  the periodic perturbation 
imposed at the entrance is carried downstream by the flow.  We also 
observe that the wave strongly evolves along the plane: in the power 
spectra figures \ref{figureevospabille}($b$) and \ref{figureevospabille} ($d$) the 
fundamental 
mode and the first harmonic of the forced wave are amplified 
downstream whereas high-frequency harmonic modes are damped.  The 
amplification of the low-frequency perturbations is also observed on 
the spectra of the natural noise.

At this stage, an important question is whether the amplification of 
the low-frequency modes of the forcing waves results from a linear 
instability or from non linear interactions between modes.  Figure 
\ref{figurecroissancelin} shows the spatial evolution of the 
fundamental mode of a forced wave along the slope for different 
forcing frequencies (the control parameters are the same as in figure 
\ref{figureevospabille}).  We observe that the forcing modes evolve 
exponentially along the plane, which implies that we are studying the 
linear development of the instability.
  
Another proof of the existence of a linear region is given in figure 
\ref{figurerelationdispersion}($a$), which shows the spatial growth 
rate $\sigma$ of a forced mode as a function of its frequency $f$.  
The black bullets are measurements obtained by imposing a forcing at 
different frequencies $f$ and measuring the growth rate of the 
fundamental mode $f$.  The other curves are obtained by keeping the 
forcing frequency $f_0$ constant and measuring the growth rate of all 
the harmonics $f_{0}$, $2f_{0}$, $3f_{0}$\ldots.  We observe that 
both 
methods give the same curves $\sigma(f)$, which shows that the 
different harmonic modes of the forced waves weakly interact and 
evolve quasi-independently in a linear regime.

  \subsection{Dispersion relation}
  

\begin{figure}
\centering
\includegraphics[scale=0.6]{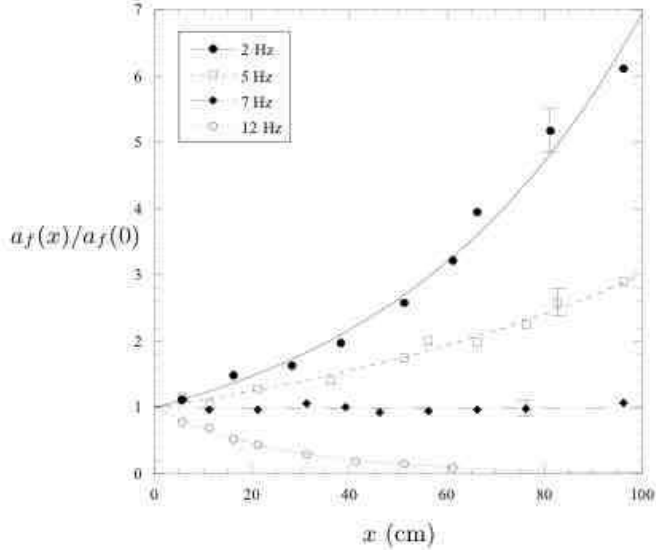}
    \caption{Spatial evolution of the amplitude of the fundamental 
    mode for different forcing frequencies. The distance $x$ is 
    measured from the nozzle. The solid curves are the best exponential fits
    $\exp (\sigma x)$ with $\sigma (2 Hz)=$0.0194 cm$^{-1}$, 
    $\sigma (5 Hz)=$0.0110 cm$^{-1}$, $\sigma (7 Hz)=$0.0001 
cm$^{-1}$ 
    and 
    $\sigma (12 Hz)=$-0.0389 cm$^{-1}$. The error bars indicate the 
    dispersion of the results. ($\theta =29^{\circ}$, $h=$ 5.3 mm). }
\label{figurecroissancelin}
\end{figure}


  \begin{figure}
	\centering
 \includegraphics[scale=0.6]{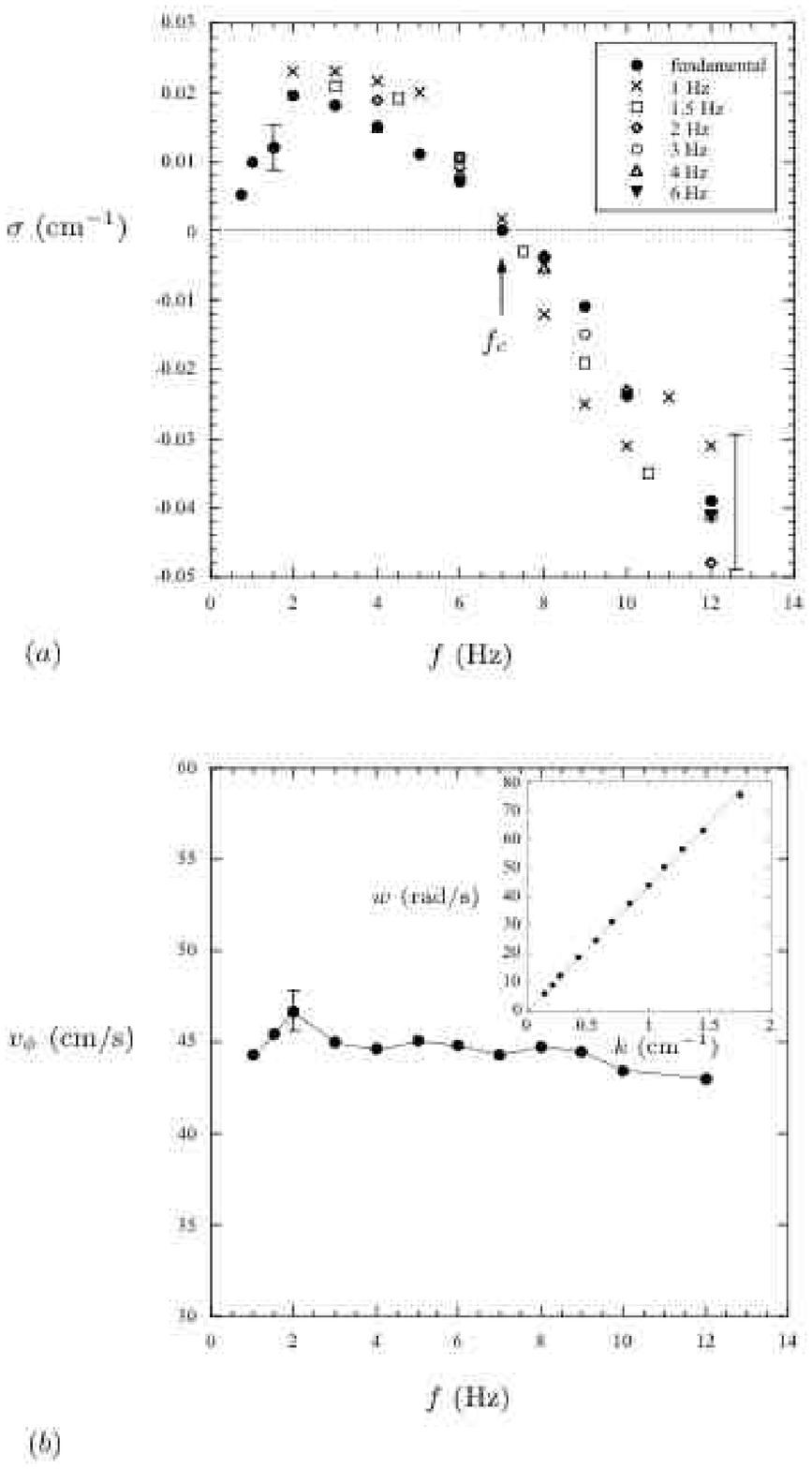} 
 \caption{Experimental dispersion relation ($\theta =29^{\circ}$, 
$h=$ 
 5.3 mm).  ($a$) Spatial growth rate $\sigma$ as a function of the 
 frequency measured from the fundamental mode of the forced wave 
 ($\bullet$) and from the harmonic modes (in that case, the forcing 
 frequency is given in the inset).  ($b$) phase velocity as a 
function 
 of the frequency.  Inset:  pulsation 
 $\omega$ as a function of the wavenumber $k$.}
\label{figurerelationdispersion}
\end{figure}

The existence of a linear region for the instability allows 
the precise measurement of the linear dispersion relation of the 
surface waves.  To measure the growth rate, the two photodetectors 
are 
separated by 60 cm.
 
A typical experimental dispersion relation is presented in figure 
\ref{figurerelationdispersion} for the above control parameters 
($\theta =29^{\circ}$, $h=$ 5.3 mm).  First, we note in figure 
\ref{figurerelationdispersion}($a$) that the spatial growth rate of a 
mode varies with the frequency.  For low-frequencies, the spatial 
growth rate is positive, i.e.  the amplitude of the mode is amplified 
along the plane as shown in figure \ref{figurecroissancelin} (2 Hz 
and 
5 Hz).  The flow is therefore unstable for the considered control 
parameters.  There exists a cutoff frequency for which the amplitude 
of the mode remains constant all along the plane (see the mode $f=$7 
Hz in figure \ref{figurecroissancelin}).  For higher frequencies, the 
amplitude decays along the plane and the mode is stable (see the mode 
$f=$12 Hz in figure \ref{figurecroissancelin}).  We shall see that 
the 
existence of a well-defined cutoff frequency $f_{c}$ will allow us to 
determine precisely the stability threshold.  The cutoff frequency 
$f_{c}$ can typically be determined with a precision of about 0.5 Hz.

The phase velocity of the wave as a function of the frequency can 
also 
be determined as shown in figure 
\ref{figurerelationdispersion}($b$).  
We observe that the phase velocity 
$v_{\phi}$ is almost independant of the frequency (the variations of 
$v_{\phi}$
 are within the error bars), which means that the system 
is non dispersive within the range of 
frequency explored in the experiment (see also the inset).  
For $\theta =29^{\circ}$ and $h=$ 5.3 mm, $v_{\phi}\approx$45 cm$/$s, 
which is a little more than the double of the average velocity of the 
flow ($u\approx$22 cm$/$s).  We have systematically measured the 
phase 
velocity for other values of the control parameters ($\theta$,$h$).  
The phase velocity is always nearly independent of the frequency.  
However, its value depends on the parameters ($\theta$,$h$) as we 
shall see in $\S\,5.4$.

\subsection{Stability threshold}

\begin{figure}
	\centering
\includegraphics[scale=0.5]{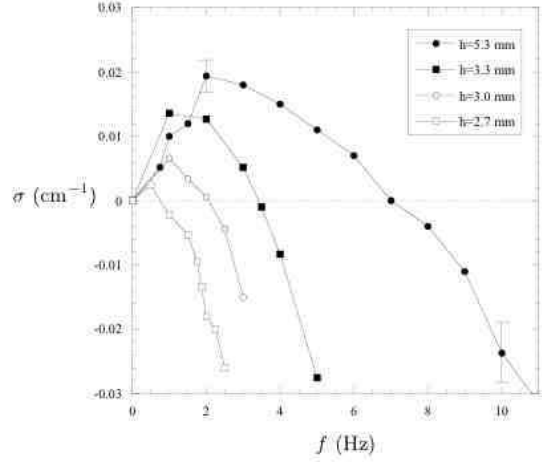}
    \caption{Spatial growth rate as a function of the frequency for 
    different thicknesses of the flow $h$ and a 
    fixed angle $\theta =29^{\circ}$. The corresponding Froude 
numbers 
    are  1.02 ($\bullet$), 0.66 
   ({\scriptsize $\blacksquare$}), 0.58 ($\circ$) et 0.51 ($\Box$).}
\label{figurerdh}
\end{figure}

So far, the results have been presented for a given inclination 
$\theta$ and thickness of the flow $h$.  When decreasing the 
thickness 
of the flow at fixed $\theta$, we observe that both the cutoff 
frequency and the growth rate of the most unstable mode decrease as 
shown in figure \ref{figurerdh}.  The flow is then less unstable as 
the thickness of the flow decreases and eventually becomes stable 
below a critical thickness $h_{c}$.

\begin{figure}
	\centering
\includegraphics[scale=0.6]{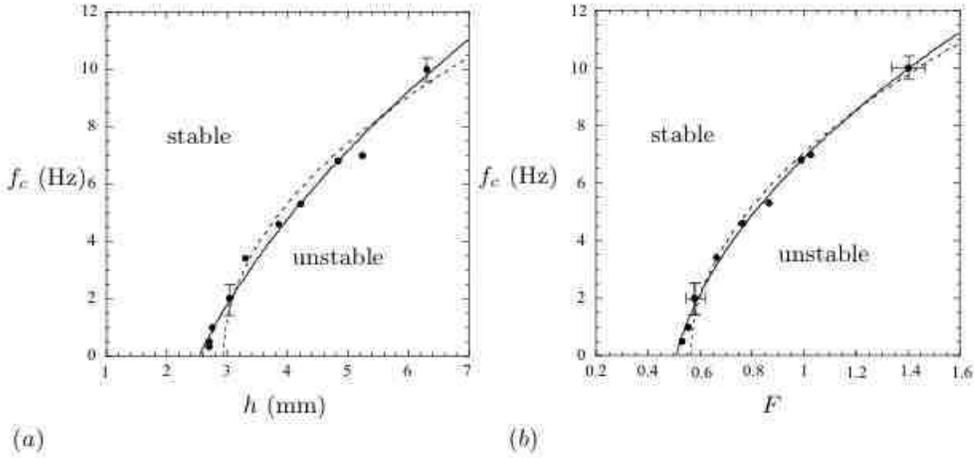}
    \caption{Neutral stability curve for $\theta 
 =29^{\circ}$. The cutoff frequency  $f_{c}$ is shown either as a 
function 
 of the thickness of the flow $h$ ($a$) or as a function of the 
Froude 
 number $F$ ($b$).
 The solid curve is a polynomial fit of the type $F, h=a 
 +bf_{c} +cf_{c}^{2}$ and the dashed curve is a quadratic fit of the 
 type $F, h=a + cf_{c}^{2}$.}
\label{figuremesureseuil}
\end{figure}

In order to measure precisely the stability threshold $h_{c}$, we 
have 
systematically measured the cutoff frequency $f_{c}$ as a function of 
the thickness $h$ as shown in figure \ref{figuremesureseuil}($a$).  
The neutral stability curve $f_{c}(h)$ is then extrapolated to zero 
frequency in order to determinate $h_{c}$.  In order to get a good 
estimation of the critical thickness we have chosen to interpolate 
the 
stability curve $f_{c}(h)$ with two polynomials (see the caption of 
figure \ref{figuremesureseuil}).  For $\theta =29^{\circ}$, we obtain 
$h_{c}(29^{\circ})=2.6 \pm$0.3 mm.  The same method can be applied to 
measure the critical Froude number $F_{c}$, where the Froude number, 
$F=u/\sqrt{gh\cos\theta}$, is measured for each set of ($\theta$,$h$)
(see figure 
\ref{figuremesureseuil}$b$).  In that case, the two control 
parameters 
are ($\theta$,$F$) instead of ($\theta$,$h$).  For $\theta 
=29^{\circ}$, we find $F_{c}(29^{\circ})=0.54 \pm$0.02.

\begin{figure}
	\centering
\includegraphics[scale=0.55]{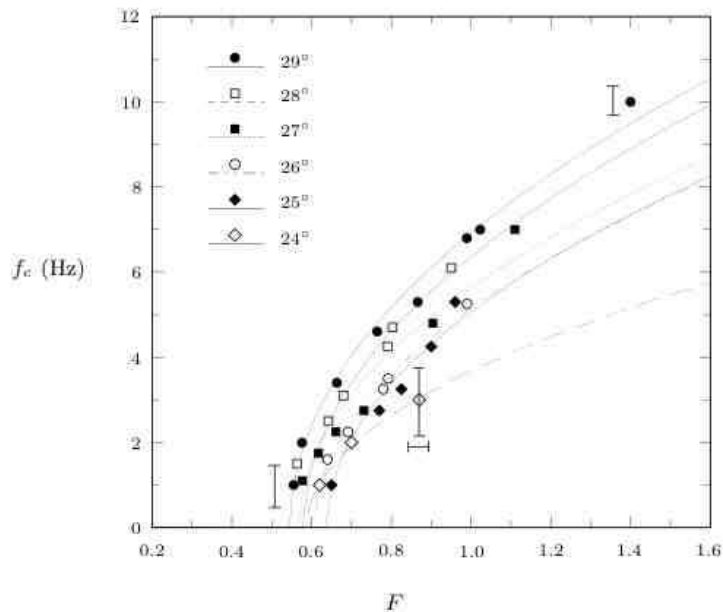}
    \caption{Neutral stability curve   $f_{c}(F)$ 
    for different inclination $\theta$. The continuous curves shows 
for each angle a fit of the type $F=a+bf_{c}^2$. }
\label{figurefctout}
\end{figure}

The entire procedure is applied for different angles of inclination.  
Results are shown in figure \ref{figurefctout} which gives the cutoff 
frequency of the waves as a function of the Froude number $F$.
  Note 
that no measurement was made for 
$\theta$ above $29^{\circ}$ or $\theta$ below $24^{\circ}$.  For 
$\theta > 29^{\circ}$, the flow is no longer in a steady uniform 
regime 
but accelerates along the slope and leaves the dense regime.
For  $\theta < 
24^{\circ}$, the flow rate required to reach the instability 
increases dramatically. Even with an amount of 
granular material as large as 150 liters, the total duration of the 
flow at the threshold is not long enough to allow precise 
measurements of the stability threshold. This limitation also 
explains the large uncertainty at $\theta = 24^{\circ}$  and why no 
measurement has been made for high Froude numbers at this angle.

We observe in figure  \ref{figurefctout} that the cutoff frequencies 
obtained at different angles 
collapse close to the threshold when plotted as a function of the 
Froude number. This result 
implies that the instability is controlled by the Froude number, at 
least near the stability threshold.

From these measurements we can determine the stability diagram of the 
flow in the phase space ($\theta$,$h$) or ($\theta$,$F$), as 
presented 
in figure \ref{figurediagstabbille}.  We note that the instability 
takes place for high inclinations and high thicknesses.  The control 
parameter is the Froude number as shown in figure 
\ref{figurediagstabbille}($b$).  While the critical thickness $h_{c}$ 
strongly varies with the angle of inclination of the plane, the 
critical Froude number $F_{c}$ weakly depends upon $\theta$.  For 
glass beads, we find $F_{c}=0.57\pm0.05$.


\subsection{Comparison with the theory}

The main experimental results of 
the instability for glass beads are: 
\begin{itemize}
\item The stability threshold is controlled by the Froude number. 
Waves  appear above a critical Froude number given by $F_{c}=0.57\pm 
0.05$. 
\item The surface wave instability is a long-wave 
instability, 
i.e.  the first unstable mode is at zero frequency (zero wavenumber).  
Above the stability threshold, there exists a cutoff frequency for 
the instability.
\item Within the frequency range investigated, the 
phase velocity of the waves weakly depends upon the frequency, i.e.  
the media is non dispersive.
\end{itemize}


In this section, we compare these results with the prediction of the 
linear stability analysis performed in the framework of the 
depth-averaged equations (see \S\,4).  We have seen that this 
analysis 
gives the stability threshold and the spatial relation of dispersion 
as a function of the velocity law $u(h,\theta)$ which, for glass 
beads, is given by (\ref{equbille}).

The only parameter that is unknown in the theory is the parameter 
$\alpha$ in the acceleration term of the Saint-Venant equations.
This parameter
is related to the unknown velocity profile across the layer (see \S\, 
4.1).  For a granular flow down an inclined plane, there is no 
experimental measurement of the velocity profile but $\alpha$ is 
close to one.  In the following, we will take $\alpha=1$ for the sake 
of 
simplicity.  We will discuss in more details the influence of 
$\alpha$ 
at the end of the section.

\subsubsection{Stability threshold}

\begin{figure}
	\centering
\includegraphics[scale=0.65]{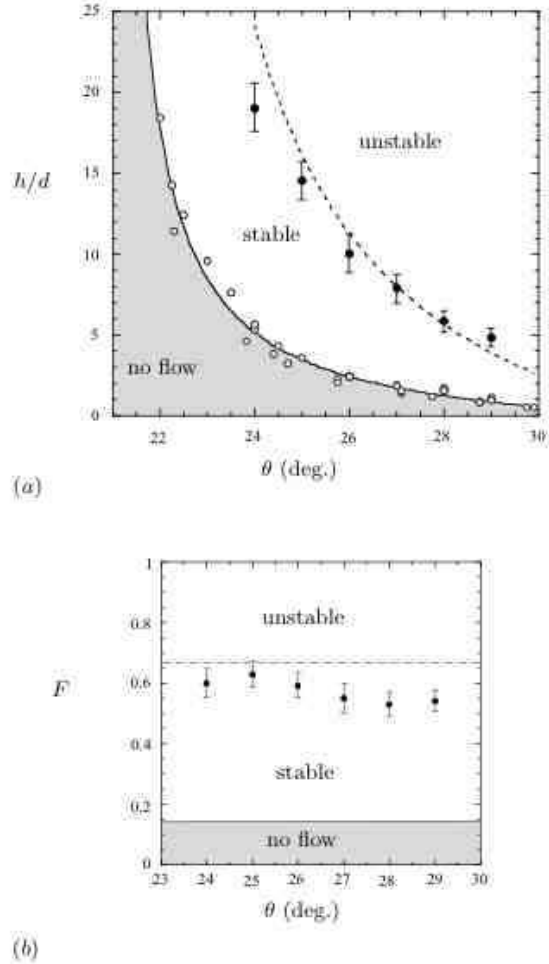}
    \caption{Stability diagram for glass beads in the phase space 
($\theta$, 
    $h$) ($a$) and ($\theta$, $F$) ($b$). 
    ($\bullet$): experiments, dashed lines:    
    theoretical predictions. The circles ($\circ$) in ($a$) is the 
deposit function $h_{stop}$.  }
\label{figurediagstabbille}
\end{figure}

We first compare the theoretical and the experimental stability 
threshold.  For glass beads, we have seen that the linear stability 
analysis predicts an instability when:
\begin{eqnarray}
	F & > & \frac{2}{3}, \label{eqseuilfroudetheobille} \\
	\mathrm{or} \;\; h & > & \frac{2}{3}\frac{\sqrt{\cos 
	\theta}}{\beta} h_{stop}(\theta).\label{eqseuilhtheobille}
\end{eqnarray}
The theoretical prediction is given in figure 
\ref{figurediagstabbille} by the dashed curve.  We note that the 
agreement between the experimental data and the theory is relatively 
good concerning the variation of the critical thickness with the 
angle 
and for the order of magnitude of the critical Froude number.  
However, the theory predicts a critical Froude number that is about 
$15-20\%$ higher than the experimental one.


\subsubsection{Dispersion relation}

\begin{figure}
	\centering
\includegraphics[scale=0.58]{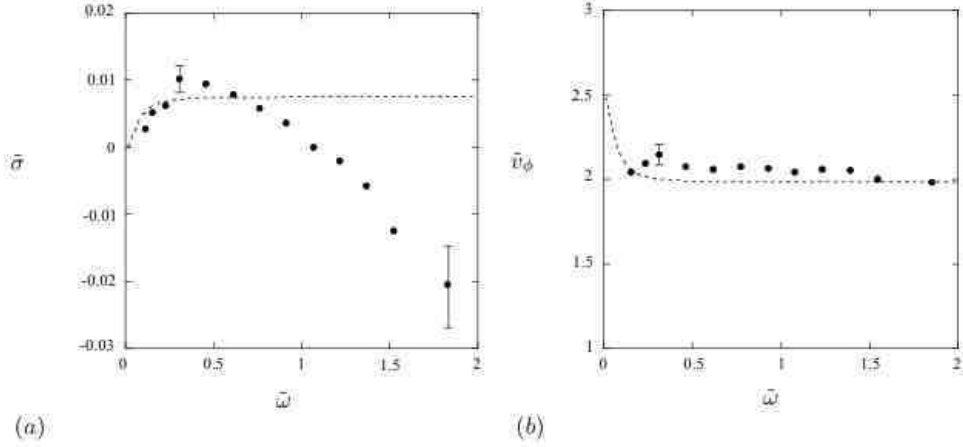}
    \caption{Theoretical (dashed line) and experimental 
    ($\bullet$) dispersion relation for 
    $\theta=29^{\circ}$ and $F=$1.02. 
    ($a$) spatial growth rate $\tilde{\sigma}$ as a function the pulsation 
    $\tilde{\omega}$. ($b$) phase velocity $\tilde{v_{\phi}}$. $\tilde{\sigma}$, $\tilde{v}_{\phi}$ 
    and $\tilde{\omega}$ are nondimensionalized using the 
    thickness of the flow $h=5.3$ mm and the averaged velocity of the 
    flow $u=$21.7 cm$/$s.}
\label{figurerdexptheo}
\end{figure}

The linear stability analysis gives also the spatial dispersion 
relation for the waves.  A typical comparison between theory and 
experiment is given in figure \ref{figurerdexptheo}, which shows on 
the same graph the predicted dispersion relation and the experimental 
data for a given set of control parameters ($\theta =29^{\circ}$, 
$F=$1.02).  We note that the main discrepancy between theory and 
experiment is that the linear stability analysis does not predict a 
cutoff frequency for the waves (figure \ref{figurerdexptheo}$a$).  No 
term in the Saint-Venant equations stabilizes the short wavelengths.  
However, the linear stability analysis gives the good order of 
magnitude for the maximum growth rate of the instability, when one 
compares the maximal growth rate measured in the experiment  and the 
maximal growth rate predicted by the theory.

  \begin{figure}
	\centering
\includegraphics[scale=0.5]{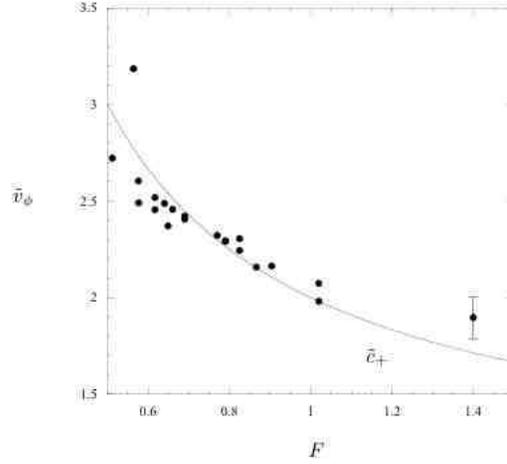}
    \caption{($\bullet$) experimental phase velocity $\tilde{v}_{\phi}$ as a 
function of the Froude 
    number $F$ for different angles of inclination 
    $(24^{\circ}-29^{\circ})$. For each ($\theta$, $F$) parameter, $\tilde{v}_{\phi}$ is obtained by averaging the dispersion relation over the frequency. The solid 
curve 
    is the velocity of the gravity waves  
     $\tilde{c}_{+}=1+(1/F)$ ($\alpha=1$).}
\label{figurevphi}
\end{figure}

It is also interesting to compare the experimental phase velocity to 
the one predicted by the linear stability analysis (figure 
\ref{figurerdexptheo}$b$).  We observe that within the frequency 
range 
investigated in the experiment, the theory predicts a constant phase 
velocity, in good agreement with the experimental data.  The theory 
also predicts an increase of the phase velocity for very low 
frequencies 
which we are unable to test with our forcing method.  It must be 
noticed that the two limits of the theoretical phase velocity have a 
precise physical meaning.  For $\tilde{\omega} = 0$, 
$\tilde{v}_{\phi}$ is equal to 
the velocity of the kinematic waves $\tilde{c}_{0}=5/2$ while for 
$\tilde{\omega}\to\infty$, $\tilde{v}_{\phi}$ tends to the velocity 
of the gravity 
waves $\tilde{c}_{+}=1+(1/F)$ (see \S\, 4.3).

We have confirmed this result by systematically measuring the phase 
velocity of the waves for different angles of inclination and Froude 
numbers.  The data are presented in figure \ref{figurevphi}.  We 
observe that the experimental phase velocity is always close to the 
speed of the gravity waves (solid line).

\subsubsection{Influence of the parameter $\alpha$ related to the 
velocity profile}

The results of the linear stability analysis presented previously are 
obtained with the parameter $\alpha$ equal to 1.  This parameter 
appears in the Saint-Venant equations in the acceleration (\S\,4.1) 
and is defined by $\alpha = \langle u^2\rangle /{\langle u\rangle 
}^2$.  It is therefore necessary to make an assumption on the 
velocity 
profile across the layer to know the value of $\alpha$. The value 
$\alpha=1$, 
corresponding to a plug flow, was used in the pioneered work of 
Savage $\&$ Hutter (1989). 
More recent studies on granular surface flows assume that the 
velocity profile is linear and use therefore the value $\alpha=4/3$ 
(Khakar  {\it et al.} 1997, Douady {\it et al.} 1999). 
For a granular flow down a rough inclined plane, recent numerical 
simulations suggest that the velocity profile does not remain linear 
for thick layers but is close to 
a Bagnold profile (\cite{ertas01}), i.e.  varies as $z^{3/2}$ ($\alpha=5/4$).
From the instability criterion (\ref{eqseuilcine}), we obtain for 
$\alpha=5/4$ (resp. $\alpha=4/3$ ) a critical Froude number 
$F_{c}=0.89$ (resp. $F_{c}=1,03$) to be compared 
with 
the experimental value $F_{c}=0.57 \pm 0.05$.  Paradoxically, the 
prediction of the theory (for both the threshold and the wave 
velocity) seems to be better with $\alpha=1$.

These results show that taking into account the velocity profile by a 
constant parameter $\alpha$ in the depth-averaged equations is not 
very satisfying.  This result is well known in the case of viscous 
liquid films.  It has been shown (see \cite{ruyer00} for instance) 
that 
the simple Saint-Venant equations do not predict quantitatively the 
primary instability if one introduces $\alpha=6/5$ 
corresponding to the parabolic velocity profile observed in viscous 
flows.  Comparison between the full linear stability analysis of the 
Navier-Stokes equations and the prediction of the Saint-Venant 
equations has shown that the latter overestimates the stability 
threshold by about 20$\%$.  This
difference between the exact three-dimensional resolution and the 
averaged equations comes from the fact that for non stationary flows, 
the shear stress at the base is no longer exactly given by its 
expression for steady uniform flows (\cite{ruyer00}).

It is therefore doubtful to assign a precise physical meaning to the 
value of the parameter $\alpha$ in the depth-averaged equations.  At 
best, one may consider $\alpha$ as a ``fit parameter'' in the 
equations and the simple value $\alpha=1$ works well in our case.

\section{Results for sand}

The visual preliminary observations with sand were different than 
with 
glass beads.  Waves were observed without forcing for slow flows but 
seemed 
to disappear for rapid flows in apparent contradiction with an 
inertial instability.  In this section, we show that the forcing 
method allows to clarify the situation.

\subsection{Experimental results}

We have measured the dispersion relation of the waves for the sand 
with the same method used for the glass beads.  However,  important difficulties arise in the case of sand. Since the flow with sand is  strongly  unstable,  it is difficult to define a linear region where the waves grow exponentially. Moreover, the large noise associated with the natural instability makes the measurements much less reproducible than with glass beads. It is therefore more difficult to quantitatively measure the dispersion relation with the sand than with glass beads. In order to measure the growth rate, we have chosen the following method. First, the  two photodiodes are located very close to the nozzle $x_{1}=2$ cm and $x_{2}=10$ cm to prevent as much as possible the non linear evolution of the waves. Then, for each frequencies $f$,  the growth rate defined as $\sigma(f)=(1/( x_{2}-x_{1})) \ln (a_{2}/a_{1})$ is averaged over many measurements, carried out at different forcing amplitudes and forcing frequencies.

\begin{figure}
	\centering
\includegraphics[scale=0.6]{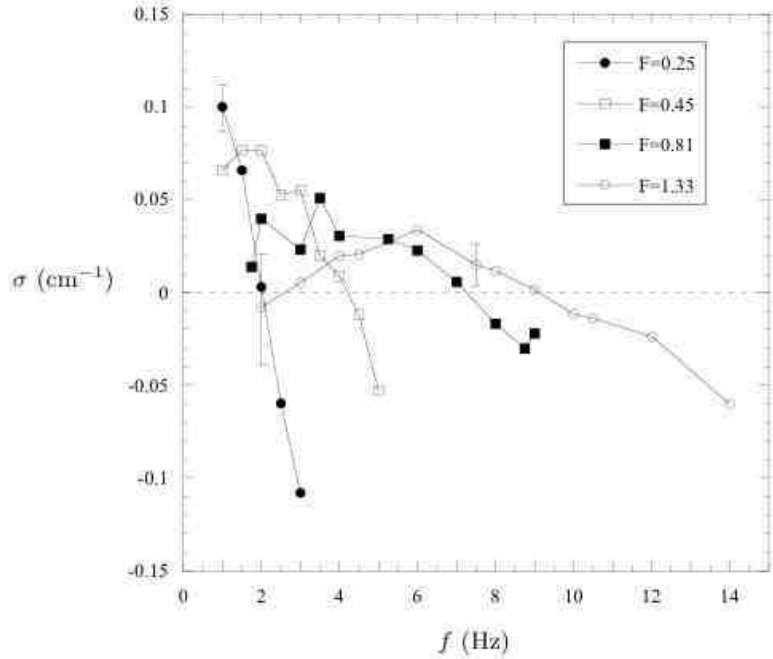}
    \caption{Results for sand 0.8 mm in diameter. Spatial growth rate as a function of frequency for 
    different Froude numbers  ($\theta =35^{\circ}$).
   The corresponding thicknesses of the flow are $h=3.3$ mm 
($\bullet$), $h=4$ mm 
   ($\Box$), $h=5$ mm 
   ({\scriptsize $\blacksquare$}) and $h=6$ mm ($\circ$).}
\label{figuresablerdF}
\end{figure}
 
In spite of the uncertainties and the large error bars, the general trend of the 
dispersion relation can be measured as shown in figure 
\ref{figuresablerdF}, which presents the evolution of the growth rate 
$\sigma(f)$ with the frequency for different Froude numbers, at a 
fixed 
angle $\theta=35^{\circ}$.  This plot has to be compared with the one 
obtained for glass beads (figure \ref{figurerdh}).  We observe that 
for 
a given Froude number, the variation of the growth rate with the 
frequency is similar to the glass beads.  The low frequencies are 
amplified while modes above a given cutoff frequency are damped.  
However, the order of magnitude of the growth rate for the sand is 
between $0.02$ cm$^{-1}$ and $0.1$ cm$^{-1}$, which is about five 
times the typical growth rate measured with glass beads.  This 
explains why the natural instability is easily observed with sand and 
not with glass beads.

Another difference between sand and glass beads arises 
 when studying the influence of the Froude 
number.  
For sand, the growth rate of the most unstable mode {\it increases} 
as 
the Froude number {\it decreases} by contrast with glass beads.  The 
most unstable mode is observed for $F=0.25$, which corresponds to the 
slowest flow that may be achieved at this angle.  This result is all 
the more surprising since the cutoff frequency decreases when the 
Froude number decreases, as observed before with glass beads.

\begin{figure}
	\centering
\includegraphics[scale=0.7]{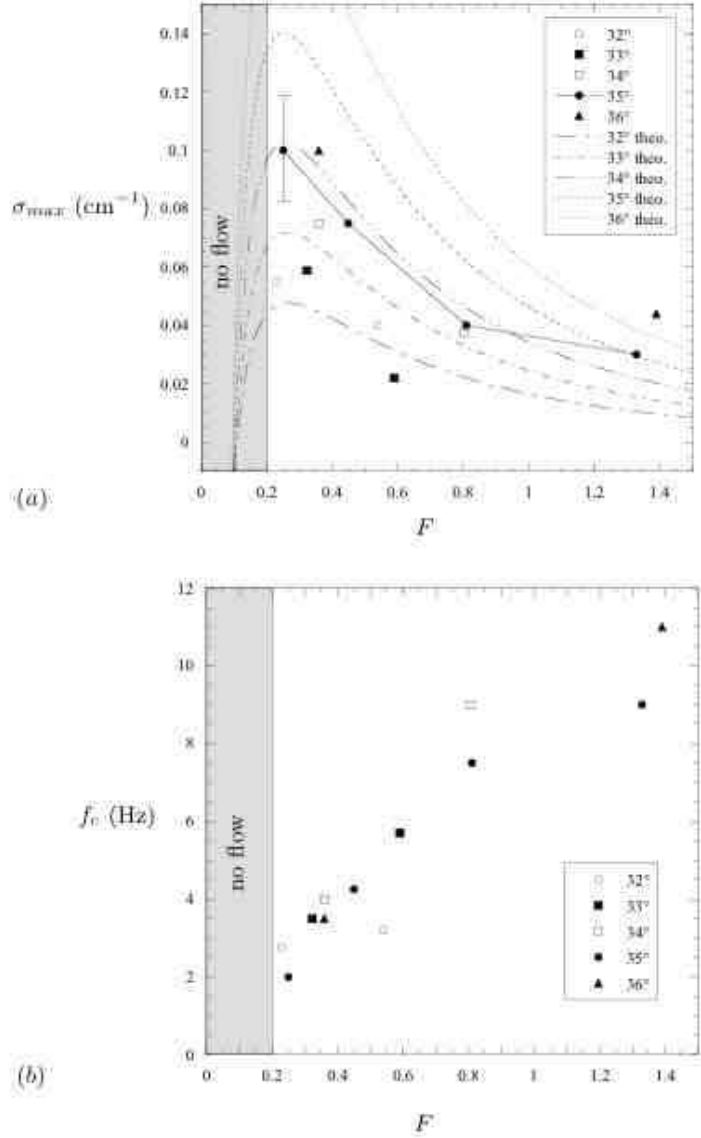}
    \caption{Results for sand 0.8mm in diameter. ($a$) spatial growth rate of the most unstable mode $\sigma_{max}(F)$; ($b$)  
    neutral stability curve $f_{c}(F)$ for different 
inclination 
    angles (given in the inset). The lines  give the 
    theoretical predictions ($\alpha=1$).}
\label{figuresabletout}
\end{figure}

 These results are confirmed for other angles of inclination.
 Figure \ref{figuresabletout}($a$) gives the growth rate of the most 
unstable mode $\sigma_{max}$ as a function of the Froude number for 
all the angles studied ($32^{\circ}<\theta<36^{\circ}$).
Considering the experimental difficulties, precise 
measurements are carried out for each angle only for two extreme 
flows: a slow flow very 
close to the flow 
threshold and a fast flow, corresponding to the limits of our set-up.
In this figure, 
we notice 
that $\sigma_{max}$ is much larger for low Froude flows close
to the flow limit (grey zone) than for high 
Froude flows.  On the other hand, 
$\sigma_{max}$ remains positive even at high Froude number, which 
means that the flow is still unstable. Therefore, the 
preliminary observation that the waves disappear at high Froude 
number 
comes simply from the fact that the maximum growth rate is a 
decreasing function of the Froude number.

It is also interesting to observe the behavior of the cutoff 
frequency $f_{c}$ with the Froude number (figure 
\ref{figuresabletout}$b$).  We notice that $f_{c}$ decreases when $F$ 
decreases, which is the signature of a 
``high-Froude-number-instability''.  However, a range 
of unstable frequencies still exists for the smallest Froude number 
$F_{min}\approx 0.25$ that can be achieved in the experiment. 
This means that, for sand, the stability threshold is pushed away below
the onset 
of flow.

\subsection{Comparison with the theory}

From the measurement presented in $\S\,4.2$ we know that the velocity 
law for flows of sand has the same analytical form 
than the velocity law for glass beads 
but with different 
coefficients.  The mean velocity $u$, the thickness $h$ and the 
inclination $\theta$ are related trough the following relation:

 \begin{equation}
	\frac{u}{\sqrt{gh}} = -\gamma + \beta \frac{h}{h_{stop}(\theta)},
	\label{eqsablerheologie}
 \end{equation}
with $\gamma = 
0.77$ and $\beta = 0.65$. 

From equation (\ref{edfrcbille}) it comes that the linear theory 
predicts an instability above the critical Froude given by:
\begin{equation}
	F_{c}=\frac{2}{3}\left( 
	1-\frac{0.77}{\sqrt{\cos\theta}}\right).
	\label{eqseuilsable}
\end{equation}
In the range of inclination we used, the predicted threshold is then 
of the order of 
$F_{c}\approx 0.1$. 

The stability threshold predicted by the linear stability analysis of 
the Saint-Venant equation for the sand is therefore close to zero and 
far below the smallest Froude number that is achieved in the 
experiment.  This prediction is compatible with our experimental 
measurements 
(figure \ref{figuresabletout}$b$) showing that the flows observed 
with 
sand are unstable from the very onset of the flow.

Concerning the dispersion relation, the theory does not describe the 
observed stabilization of the short wavelengths, as already noticed 
for 
glass beads.  No cutoff frequency is predicted.  However, we can 
compute for a given set of control parameters ($\theta$, $F$) the 
maximum spatial growth rate $\sigma_{max}$.  In the theory, the 
maximum growth rate is achieved at infinite frequency.  Using 
(\ref{eqsablerheologie}) for the law 
$u(h,\theta)$  and the expression of the 
spatial growth rate (\ref{eqsigmamaxa}) from Appendix A, we 
obtain ($\alpha=1$):
\begin{equation}
\sigma_{max} = -\frac{1}{h}\tilde{k}_{i}(+\infty) = \frac{a}{h} 
\left[\frac{
\tilde{c}_{0}-1-(\frac{1}{F})}{2(F+1)}\right],
\label{eqsigmamaxsable}
\end{equation}
where $a$ is given by (\ref{eqabgranul}) and $\tilde{c}_{0}$ is given by (\ref{eqc0granul}).

In figure \ref{figuresabletout}($a$) the lines give the prediction 
for $\sigma_{max}(F, \theta)$.  We first notice that in the range of 
Froude number where measurements are carried out, the predicted 
maximum 
growth rate is always positive, i.e.  the flow is always unstable.  
Secondly, for $F\gtrsim 0.25$, the maximal growth rate predicted by 
the theory decreases when the Froude number increases, as observed in 
the experiment.  Finally, although there is no quantitative agreement between the predicted maximal growth rates and those measured experimentally, the order of magnitude of    $\sigma_{max}$ predicted by the theory is of the same as the order of magnitude as the experimental measurements.

\subsection{Discussion}

The linear stability analysis of the Saint-Venant equations seems 
therefore sufficient to understand the properties of sand flow that 
were first surprising in the observation of the unforced system.  The 
theory predicts that the flow is always unstable and that the most 
unstable modes occur for very slow flows, close to the flow 
threshold.  
Therefore the occurrence of waves for slow flows in sand does not 
come 
from a new instability mechanism but results from the classical long-wave
 inertial instability.

However, the features of the instability contrast dramatically with 
the instability in classical fluids due to the difference in rheology 
and flow rules.  
A granular media can be unstable from the very onset of the flow 
unlike classical fluid flows.  This property actually results from 
the 
existence of a critical angle in granular flows.  Unlike classical 
fluids, the speed of the kinematic waves in a granular flow does not 
vanish at the flow threshold, i.e.  when $h\to h_{stop}$.  If the 
speed of the kinematic wave $c_{0}$ at $h\to h_{stop}$, 
$c_{0}=u(h_{stop})+{u'\vert}_{(h_{stop})}\,h_{stop}$, is larger than 
the speed 
$c_{+}$ of 
the gravity waves, $c_{+}=u(h_{stop})+\sqrt{gh_{stop}\cos\theta}$, 
the flow may be 
unstable from the very onset of the flow.  This explains why the 
Saint-Venant equations can predict the formation of waves with an 
inertial mechanism even when the mean velocity goes to zero.


\begin{figure}
	\centering
\includegraphics[scale=0.6]{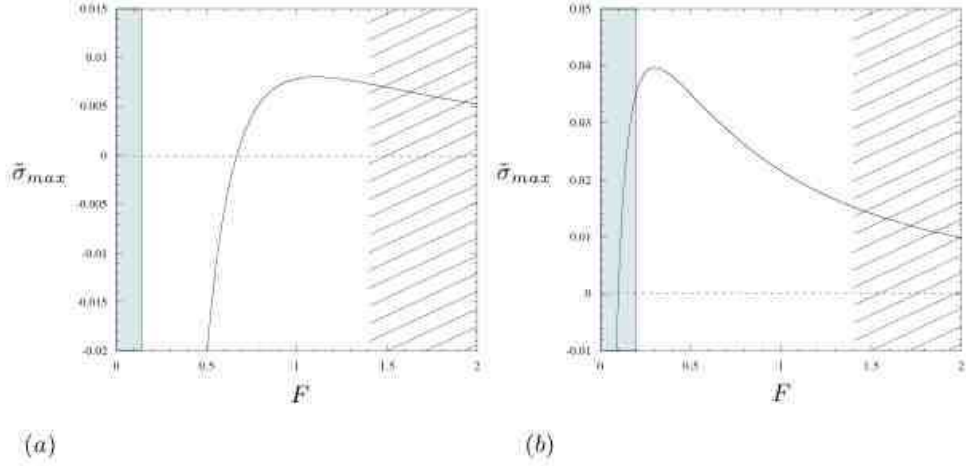}
    \caption{Predicted dimensionless growth rate of the most unstable mode as a function of 
    the Froude number for glass beads ($\theta=29^{\circ}$) ($a$) and sand ($\theta=35^{\circ}$) ($b$) ($\alpha=1$). Grey zone: no flow. Hatched zone: no measurements. }
\label{figurediscussion}
\end{figure}

Furthermore, the difference between glass beads and sand becomes 
clear in the 
light of the linear stability analysis.  The instability mechanism is 
the same. However, because of quantitative differences in the 
coefficients of the flow rule, the characteristics of the instability 
differ in the range where experimental measurements are possible.  
This is clearly shown in figure \ref{figurediscussion} where the 
maximal growth rate predicted by the theory is plotted as a function 
of the Froude number for both glass beads and sand.  We also indicate 
the range of Froude number where experiments are carried out.  It is 
clear from figure \ref{figurediscussion} that there is no qualitative 
difference between both systems.  The difference lies only in the 
relative position of the unstable region compared to the measurement 
region.  For beads the flow threshold is below the instability 
threshold and the measurements are carried out in a region where the 
maximum growth rate mainly increases with the Froude number.  By 
contrast, 
for sand the flow threshold is above the instability threshold and 
the 
measurements are carried out in a region where the growth rate 
decreases with the Froude number.

\section{Conclusion}

We have presented in this paper an experimental study of the long 
surface waves instability for dense granular flows down rough 
inclined 
planes.  By imposing a controlled perturbation at the entrance of the 
flow, we have been able to precisely measure the threshold and the 
dispersion relation of the instability.

Using glass beads we have shown that the long-wave instability is 
controlled by the Froude number and occurs above a critical Froude 
number.  The stability threshold and the velocity of the waves 
measured experimentally are quantitatively in good agreement with the 
predictions of a linear stability analysis of the Saint-Venant 
equations.  Using sand, we have observed that the properties of the 
long-wave instability are strongly modified: the flow is always 
unstable and the most unstable modes occur for slow flows, near the 
onset of the flow.  Despite the apparent qualitative differences 
between both systems, we have shown that the wave formation in both 
cases results from the same instability mechanism and can be 
described 
by the stability analysis of the Saint-Venant equations.  The 
difference between sand and glass beads only comes from quantitative 
differences in the coefficient of the friction law introduced in the 
Saint-Venant equations.

The instability mechanism for the waves formation in granular flows 
therefore results, as for classical fluids, from the competition 
between inertia and gravity.  However, the features of the 
``classical'' long-wave instability can be modified due to the 
specificity of the friction law for granular flows.  The most 
dramatic 
effect is that the stability threshold may be lowered below the onset 
of the flow, i.e.  a granular flow may be unstable as soon as it 
flows, when the velocity is small.  This property strongly 
distinguishes 
granular flows from classical fluid flows and is closely related to 
the 
existence of a critical angle in granular material.  The existence of 
a strong instability close to the onset of the flow certainly has 
dramatic 
consequences for the non linear evolution of the waves.  When the 
deformations develop, the thickness of the layer can rapidly become 
less than the minimum thickness needed to flow.  
The flow then evolves towards a succession of surges separated by 
material at rest.  A similar behavior may be expected for other 
materials as soon as the material will present a yield stress (mud, 
clay, granular\ldots).  This study then suggests that the existence 
of 
surges often observed in geophysical flows could result from the 
existence of an instability at the onset of flow due to a non zero 
yield stress condition of the material.

Another interesting result of this study is the relative success of 
the depth-averaged equations in predicting the stability properties 
of 
granular flows.  As soon as a relevant friction law is taken into 
account, quantitative properties can be predicted. 
  This work then 
provides a good test for the relevance of the friction law deduced 
from the steady uniform flows, in a case where inertial effects 
determine the dynamics of the flow. 

 However, our study reveals an important limitation of the 
 depth-averaged approach we use for describing the instability.  The 
 simple first order Saint-Venant equations are unable to predict the 
 cutoff frequency 
 observed in the experiment.  This cutoff frequency is the signature 
 of a dissipative mechanism that stabilizes short wavelengths.  Since 
 dry granular flows do not experience surface tension, this 
 stabilization mechanism should be related to the streamwise velocity 
 variations that are second order effects in a shallow water 
 description.  In order to take into account these longitudinal 
 gradients, one should {\it a priori} know the full three-dimensional 
 constitutive equations, which is still an open problem for dense 
 granular flows.  Our measurement of the surface wave instability and 
 of the cutoff frequency could therefore serve as a test for future 
 propositions of constitutive equations.

 \acknowledgements
 This research was supported by the Minist\`ere Fran\c{c}ais de la 
 Recherche (ACI ``Jeunes Chercheurs'' $\# 2018$ and ``Pr\'evention des Catastrophes Naturelles''). We thank J. Vallance for 
 stimulating discussions and V. Desbost and P. Ferrero for their 
participation to 
the preliminary experiments. We thank F. Ratouchniak
for his technical 
assistance.

\appendix

\section{}

We give in this Appendix the spatial stability analysis of the 
dispersion relation (\ref{eqreldisp}) given by:
 \begin{equation}
	-\tilde{\omega}^2 + 2\alpha \tilde{\omega}\tilde{k} + 
\frac{i}{F^2}\left( (a-b)\tilde{k}-a\tilde{\omega} 
	\right) + \left( \frac{1}{F^2}-\alpha \right) \tilde{k}^2= 0.
	\label{eqreldispa}
\end{equation}
The pulsation $\tilde{\omega}$ is real and the wavenumber $\tilde{k}$ 
is 
complex: $\tilde{k}= \tilde{k}_{r}+i\tilde{k}_{i}$,  the flow is then 
unstable when 
$\tilde{k}_{i}\tilde{k}_{r}<0$. The resolution of the dispersion 
relation (\ref 
{eqreldispa}) gives two spatial modes $(+)$ and $(-)$ for 
$\tilde{k}(\tilde{\omega})$ 
that are given by:
\begin{eqnarray}
	{\tilde{k}_{r}}^{\pm} & = & \frac{\alpha}{(\alpha-\frac{1}{F^{2}})} 
	\tilde{\omega} \mp \frac{\sqrt{2}(\frac{a}{F^{2}}-\alpha 
	b)}{F^2(\alpha-\frac{1}{F^{2}}) }\tilde{\omega} {\left[ 
-g(\tilde{\omega}) + {\left( 
	{g(\tilde{\omega})}^2 +\frac{16\tilde{\omega}^2} {F^4} 
(\frac{a}{F^{2}}-\alpha b)^2 
	\right) }^{\frac{1}{2}} \right] }^{-\frac{1}{2}}\\
	{\tilde{k}_{i}}^{\pm} & = & \frac{a-b}{2 
F^2(\alpha-\frac{1}{F^{2}})} 
	\mp \frac{1}{2\sqrt{2}(\alpha-\frac{1}{F^{2}})} {\left[ 
-g(\tilde{\omega}) + 
	{\left( {g(\tilde{\omega})}^2 + \frac{16\tilde{\omega}^2} {F^4} 
	(\frac{a}{F^{2}}-\alpha b)^2 \right) }^{\frac{1}{2}} \right] 
	}^{\frac{1}{2}}
\end{eqnarray}
where $g(\tilde{\omega})$ is given by:
\begin{equation}
	g(\tilde{\omega}) = 4\left( \alpha(\alpha - 1) + 
\frac{1}{F^2}\right)\tilde{ \omega}^2 
	- \frac{(a-b)^{2}}{ F^4}.
\label{eqdefga}
\end{equation}
For the friction law considered in the article the parameter $a$ is 
positive and the parameter $b$ is negative. The 
mode $(-)$ is then always stable whereas the mode $(+)$ may be stable 
or 
unstable depending upon the Froude number or the angle of 
inclination. 
However it can be shown that the sign of 
$\tilde{k}_{i}\tilde{k}_{r}$, i.e. the 
stability of the flow, does not depends 
upon the pulsation $\tilde{\omega}$. Therefore, to find the stability 
threshold we can 
study the asymptotic form of the dispersion relation. For 
$\tilde{\omega}\to\infty$, one finds:
\begin{eqnarray}
	{\tilde{k}_{r}}^{\pm}(+\infty) & = &\frac{ \tilde{\omega} }{ \alpha 
\pm \sqrt{\alpha 
	(\alpha -1) +\frac{1}{F^{2}}} },\label{krmaxa}\\
	{\tilde{k}_{i}}^{\pm}(+\infty) & = & \frac{ \mp a\left( 
1-\frac{b}{a} - 
\alpha \mp 
	\sqrt{\alpha (\alpha - 1) + \frac{1}{F^2}}\right)} {2 F^{2} 
	\left( \alpha \pm \sqrt{\alpha (\alpha -1) +\frac{1}{F^{2}}} 
	\right) \sqrt{\alpha(\alpha -1) +\frac{1}{F^{2}} } }.
	\label{eqsigmamaxa}
\end{eqnarray}
The mode $(+)$ is then unstable when:
\begin{equation}
1-\frac{b}{a}> \alpha + \sqrt{\alpha(\alpha - 1) + \frac{1}{F^2}}.
\label{eqseuilspaa}
\end{equation}
  
\section{}

We give here another way to determine the instability condition 
(\ref{eqseuilcine}), which underlines the competition between the 
kinematic waves and the gravity waves in the stability of the flow.

 By differentiating (\ref{eqlinm}) with respect to 
$\tilde{t}$ (resp. with respect to $\tilde{x}$) and (\ref{eqlinp}) with respect to $\tilde{x}$, the 
perturbed velocity 
field $u_{1}$ may be eliminated and one obtains a partial linear 
equation for $h_{1}$ given by: 
\begin{equation}
\frac{{\partial}^2 h_{1}}{\partial \tilde{t}^2} +2\alpha 
\frac{{\partial}^2 
h_{1}}{\partial \tilde{x} \partial \tilde{t}} + (\alpha - 
\frac{1}{F^2}) 
\frac{{\partial}^2 h_{1}}{\partial \tilde{x}^2} = - 
\frac{a}{F^2}\left(\frac{{\partial} h_{1}}{\partial \tilde{t}} + 
(1-\frac{b}{a})\frac{{\partial} h_{1}}{\partial \tilde{x}}\right),
\label{eqhlin1}
\end{equation}
We then re-write (\ref{eqhlin1}) as:
\begin{equation}
	\left( \frac{\partial}{\partial \tilde{t}} + \tilde{c}_{+} 
	\frac{\partial}{\partial \tilde{x}}\right) \left( 
\frac{\partial}{\partial 
	\tilde{t}} + \tilde{c}_{-} \frac{\partial}{\partial 
\tilde{x}}\right) h_{1} = - 
	\frac{a}{F^2}\left(\frac{{\partial} h_{1}}{\partial \tilde{t}} + 
	\tilde{c}_{0}\frac{{\partial} h_{1}}{\partial \tilde{x}}\right),
\label{eqhierarchie}
\end{equation}
where $\tilde{c}_{\pm} = \alpha \pm \sqrt{\alpha(\alpha - 1) + 
(1/F^2)}$ 
is the speed of the gravity waves upstream and downstream and 
$\tilde{c}_{0}= 
1-(b/a)=
1+( \partial \tilde{u}/\partial 
\tilde{h})_{0}$ is the speed of the kinematic waves. Equation 
(\ref{eqhierarchie}) reveals  waves hierarchy in the system 
(\cite{witham74}). Long wavelength perturbations propagate to the 
first 
order as kinematic waves with a velocity $\tilde{c}_{0}$. The effect 
of 
higher order terms may be captured by substituting $\partial / 
\partial \tilde{t} \sim 
-\tilde{c}_{0} \partial / \partial \tilde{x}$ on the left-hand side 
of (\ref{eqhierarchie}), which leads to:
\begin{equation}	
(\tilde{c}_{0}-\tilde{c}_{-})(\tilde{c}_{+}-\tilde{c}_{0})\frac{{\partial}^2 
h_{1}}{\partial \tilde{x}^2} = 
	\frac{a}{F^2} \left(\frac{{\partial} h_{1}}{\partial \tilde{t}} + 
	\tilde{c}_{0}\frac{{\partial} h_{1}}{\partial \tilde{x}}\right).
\label{eqdif}
\end{equation}
This equation is a combined diffusion equation and the stability of 
the 
perturbation is given by the sign of the ``diffusion coefficient'':  
$(F^{2}/a)(\tilde{c}_{0}-\tilde{c}_{-})(\tilde{c}_{+}-\tilde{c}_{0})$. 
For the friction law considered in this paper, $a>0$ and 
$b<0$, i. e. $\tilde{c}_{0}>1$. On the other hand, $\tilde{c}_{-}<1$ since $\alpha>1$ and $F>0$. 
Therefore $(\tilde{c}_{0}-\tilde{c}_{-})$ is always positive and  the 
flow is unstable when:
\begin{equation}
	\tilde{c}_{0}>\tilde{c}_{+}.
	\label{eqseuilcinea}
\end{equation}


\end{document}